\documentclass[aps,prd,preprint,nofootinbib,amsmath,amssymb,superscriptaddress]{revtex4-1}

 \usepackage{amsmath}
\usepackage{amssymb}
\usepackage{graphicx}
\usepackage{soul,color}
\usepackage{bm}
\usepackage{times}
\usepackage{hyperref}
\hypersetup{
  colorlinks=true,
  citecolor=blue,
  linkcolor=blue,
  urlcolor=blue}

\usepackage{epsfig}
\usepackage{dcolumn}
\usepackage{float}

\usepackage{epsfig}
\usepackage{dcolumn}
\def\be{\begin{equation}}
\def\ee{\end{equation}}
\def\bea{\begin{eqnarray}}
\def\eea{\end{eqnarray}}
\def\gsim{\ \rlap{\raise 2pt\hbox{$>$}}{\lower 2pt \hbox{$\sim$}}\ }
\def\lsim{\ \rlap{\raise 2pt\hbox{$<$}}{\lower 2pt \hbox{$\sim$}}\ }
\def\dslash{\kern-4pt \not{\hbox{\kern-2pt $\partial$}}}
\def\pslash{\not{\hbox{\kern-2pt p}}}

\newcommand{\dcp}{\delta_{CP}}

\newcommand{\cnv}{\v{C}erenkov}

\begin{document}

\setstcolor{red}

\DeclareGraphicsExtensions{.eps,.ps}

\title{Testing NSI suggested by solar neutrino tension in T2HKK and DUNE}

\author{Monojit Ghosh}
\email{manojit@kth.se}
\affiliation{Department of Physics, Tokyo Metropolitan University, Hachioji,  Tokyo 192-0397, Japan}
\affiliation{Department of Physics, School of Engineering Sciences, KTH Royal Institute of Technology,\\ AlbaNova University Center, Roslagstullsbacken 21, SE--106 91 Stockholm, Sweden }
\affiliation{The Oskar Klein Centre, AlbaNova University Center, Roslagstullsbacken 21,\\ SE--106 91 Stockholm, Sweden}

\author{Osamu Yasuda}
\email{yasuda@phys.se.tmu.ac.jp}
\affiliation{Department of Physics, Tokyo Metropolitan University, Hachioji, Tokyo 192-0397, Japan}

\begin{abstract}
It was shown that the tension between the mass-squared differences  
obtained from solar neutrinos and those acquired through KamLAND experiments may be solved
by the introduction of a non-standard flavor-dependent interaction (NSI) in
neutrino propagation.
In this study, we discuss the possibility of testing
such a hypothesis using the future long-baseline neutrino experiments
T2HKK and DUNE.
Assuming that the NSI does not exist,
we provide the excluded region within the ($\epsilon_D$, $\epsilon_N$) plane,
where $\epsilon_D$ and $\epsilon_N$ are the parameters appearing
in the solar neutrino analysis conducted with the NSI.
We find that the best-fit value from the solar neutrino and KamLAND data
(global analysis of a particular coupling to quarks)
can be tested at more than 10$\sigma$ (3$\sigma$) by these two
experiments for most of the parameter space.

\end{abstract}


\maketitle

\section{Introduction}
\label{sec1}

It has been well established by solar, atmospheric, reactor and accelerator neutrino
experiments that neutrinos have mass and mixings\,\cite{Patrignani:2016xqp}.
In the standard three flavor neutrino oscillation framework,
there are three mixing angles $\theta_{12}$, $\theta_{13}$, $\theta_{23}$, 
two mass-squared differences $\Delta m^2_{31}$ and $\Delta m^2_{21}$, and one Dirac type CP phase $\delta_{CP}$.
The approximate values of the oscillation parameters are determined as follows:
$(\Delta m^2_{21},\sin^22\theta_{12}) \simeq (7.5\times 10^{-5}$eV$^2$,$0.86)$,
$(|\Delta m^2_{31}|,\sin^22\theta_{23}) \simeq (2.5\times 10^{-3}$eV$^2$,$1.0)$,
$\sin^22\theta_{13}\simeq 0.09$ \cite{Capozzi:2017ipn,deSalas:2017kay,Esteban:2016qun}.
Although there are some indications that $\delta_{\rm CP}\sim-90^\circ$
and $\Delta m^2_{31}>0$ are favored, we do not know
the value of the Dirac CP phase $\delta_{\rm CP}$, the sign of $\Delta
m^2_{31}$ (the mass hierarchy), or the octant of $\theta_{23}$ (the
sign of $45^\circ-\theta_{23}$) with a high degree of confidence.
To measure these undetermined neutrino oscillation parameters,
neutrino oscillation experiments
with high statistics, such as
T2HK\,\cite{Abe:2011ts},
DUNE\,\cite{Acciarri:2015uup},
and T2HKK\,\cite{Abe:2016ero}
have been proposed.  With these precision measurements, we
can also probe the new physics by looking at the deviation from
the standard three flavor neutrino mixing scenario.

However, it is well known\,\cite{deHolanda:2010am,Gonzalez-Garcia:2013usa}
that tension occurs between the mass-squared difference deduced from the
solar neutrino observations and that derived from the KamLAND experiment.
While Ref.\,\cite{deHolanda:2010am} proposed a sterile
neutrino oscillation with a mass-squared difference
on the order of O(10$^{-5}$) eV$^2$ as a solution of this tension,
it was pointed out in Ref.\,\cite{Gonzalez-Garcia:2013usa} 
that the tension can be resolved by introducing flavor-dependent
non-standard interactions (NSI)
in neutrino propagation:\cite{Wolfenstein:1977ue,Guzzo:1991hi,Roulet:1991sm}
\begin{eqnarray}
\hspace{-8mm}&{\ }&
{\cal L}^{\mbox{\tiny{\rm NSI}}} 
=-2\sqrt{2}\, \epsilon_{\alpha\beta}^{ff'P} G_F
\bar{\nu}_{\alpha L} \gamma_\mu \nu_{\beta L}\,
\bar{f}_P \gamma^\mu f_P',
\label{NSIop}
\end{eqnarray}
where $f_P$ and $f_P'$ are fermions with chirality $P$,
$\epsilon_{\alpha\beta}^{ff'P}$ is a dimensionless constant, and $G_F$ is the Fermi coupling constant.
Constraints on $\epsilon_{\alpha\beta}$
have been previously discussed by numerous researchers\,\footnote{
See Refs.\,\cite{Ohlsson:2012kf,Miranda:2015dra} for extensive references.},
from atmospheric neutrinos\,\cite{GonzalezGarcia:1998hj,Lipari:1999vh,Fornengo:1999zp,Fornengo:2001pm,GonzalezGarcia:2004wg},
$e^+ e^-$ colliders\,\cite{Berezhiani:2001rs},
the compilation of various neutrino data\,\cite{Davidson:2003ha,Biggio:2009nt},
solar neutrinos\,\cite{Friedland:2004pp,Miranda:2004nb,Palazzo:2009rb},
$\nu_e e$ or $\bar{\nu}_e e$ scatterings\,\cite{Barranco:2005ps,Barranco:2007ej},
solar and reactor neutrinos\,\cite{Bolanos:2008km},
and solar, reactor, and accelerator neutrinos\,\cite{Escrihuela:2009up}.
The constraints on $\epsilon_{ee}$ and $\epsilon_{e\tau}$
from atmospheric neutrino have been discussed in
Ref.~\cite{GonzalezGarcia:2011my} along with those
from long-baseline experiments,
in Ref.~\cite{Mitsuka:2011ty} by the Super-Kamiokande Collaboration,
in Refs.~\cite{Ohlsson:2013epa,Esmaili:2013fva,Chatterjee:2014gxa,Choubey:2014iia,Choubey:2015xha}
which discussed future atmospheric neutrino experiments. NSI has recently been studied extensively in terms of long-baseline
experiments \cite{Friedland:2012tq,Adhikari:2012vc,Masud:2015xva,deGouvea:2015ndi,Rahman:2015vqa,Coloma:2015kiu,Liao:2016hsa,Soumya:2016enw,Blennow:2016etl,Forero:2016cmb,Huitu:2016bmb,Bakhti:2016prn,
 Masud:2016bvp,Coloma:2016gei,Masud:2016gcl,Agarwalla:2016fkh,Ge:2016dlx,Liao:2016bgf,Fukasawa:2016gvm,Blennow:2016jkn,Liao:2016orc,Deepthi:2016erc,Fukasawa:2016lew,
 Ghosh:2017ged,Masud:2017bcf}\footnote{Apart from the NSIs occurring during neutrino propagation, NSIs also take place in neutrino production and detection. Such charge current NSIs are more relevant to low-energy experiments \cite{Khan:2016uon,Khan:2017oxw,Tang:2017qen}.}.
It is known that some models predict
large non-standard interactions\,\cite{Farzan:2015doa,Farzan:2016wym,Farzan:2016fmy},
and hence such large NSI effects are worth investigating from the viewpoint
of model building.

In the analysis of long-baseline experiments and
atmospheric neutrino experiments, the dominant
contribution comes from the larger mass squared difference
$\Delta m_{31}^2$ and the oscillation probabilities are
expressed in terms of $\epsilon_{\alpha\beta}$ and
the standard oscillation parameters.
While the results in Ref.~\cite{Gonzalez-Garcia:2013usa}
may suggest the existence of NSI,
the parametrizations for the NSI parameters
($\epsilon_D$, $\epsilon_N$)
in Ref.~\cite{Gonzalez-Garcia:2013usa} are different
from those with $\epsilon_{\alpha\beta}$, and
it remains unclear how the allowed region in Ref.~\cite{Gonzalez-Garcia:2013usa}
will be tested or excluded through future experiments.
In Ref.\,\cite{Fukasawa:2016nwn}, assuming a standard oscillation scenario,
the excluded region within the ($\epsilon_D$, $\epsilon_N$) plane
was given for the atmospheric neutrino measurements at Hyper-Kamiokande.
In this study, we discuss the sensitivity of the
accelerator-based neutrino measurements T2HKK and DUNE to NSIs
using the same parametrization as described in Ref.\,\cite{Gonzalez-Garcia:2013usa}.
Because the parametrization used in Ref.\,\cite{Gonzalez-Garcia:2013usa} differs from $\epsilon_{\alpha\beta}$ on a three flavor basis, a non-trivial mapping
is required to compare the results of these two parametrizations.

As with the case of a standard scenario\,\cite{BurguetCastell:2001ez,Minakata:2001qm,Fogli:1996pv,Barger:2001yr,Ghosh:2015ena},
parameter degeneracy in the presence of the new physics
has also been studied in Refs.\,\cite{Bakhti:2014pva,Mocioiu:2014gua,Coloma:2015kiu,Liao:2016orc,Coloma:2016gei,Blennow:2016etl,Agarwalla:2016fkh,Deepthi:2016erc,Liao:2016orc}.
Because little is known regarding parameter degeneracy in the
parametrization of $\epsilon_D$ and $\epsilon_N$,
and because such a study is beyond the scope of this paper,
we do not discuss parameter degeneracy herein.

The remainder of this paper is organized as follows.
In section \ref{sec2}, we
describe the neutrino oscillations in the presence
of NSIs in neutrino propagation, as well as descriptions of the T2HKK and DUNE experiments.
In section \ref{sec3}, 
we describe the correspondence between the
parametrization $\epsilon_{\alpha\beta}$ in the long baseline experiments
and ($\epsilon_D$, $\epsilon_N$) in the solar neutrino experiments.
In section \ref{sec4},
we describe our results.
In section \ref{sec5}, we draw our concluding remarks.

\section{Three flavor neutrino oscillation framework with NSI\label{3nu_nsi}}
\label{sec2}
\subsection{Nonstandard interactions\label{nsi_intro}}
The presence of NSIs (Eq. (\ref{NSIop})) modifies the neutrino evolution
governed by the positive energy part of the Dirac equation:
\begin{eqnarray}
i {d \over dx} \left( \begin{array}{c} \nu_e(x) \\ \nu_{\mu}(x) \\ 
\nu_{\tau}(x)
\end{array} \right)
= \left\{
U \,{\rm diag} \left(0, \Delta E_{21}, \Delta E_{31}
\right)  U^{-1}
+{\cal A}\right\}
\left( \begin{array}{c}
\nu_e(x) \\ \nu_{\mu}(x) \\ \nu_{\tau}(x)
\end{array} \right)\,,
\label{eqn:sch}
\end{eqnarray}
where
$U$ is the leptonic mixing matrix defined by
\begin{widetext}
\begin{eqnarray}
&{\ }&\hspace{-2mm}
U=
\left(
\begin{array}{ccc}
c_{12}c_{13} & s_{12}c_{13} &  s_{13}e^{-i\dcp}\\
-s_{12}c_{23}-c_{12}s_{23}s_{13}e^{i\dcp} & 
c_{12}c_{23}-s_{12}s_{23}s_{13}e^{i\dcp} & s_{23}c_{13}\\
s_{12}s_{23}-c_{12}c_{23}s_{13}e^{i\dcp} & 
-c_{12}s_{23}-s_{12}c_{23}s_{13}e^{i\dcp} & c_{23}c_{13}
\end{array}
\right)\,,
\label{eqn:mns3}
\end{eqnarray}
\end{widetext}
and $\Delta E_{jk}\equiv\Delta m_{jk}^2/2E\equiv (m_j^2-m_k^2)/2E$,
$c_{jk}\equiv\cos\theta_{jk}$, and $s_{jk}\equiv\sin\theta_{jk}$.
In addition, ${\cal A}$ in Eq.\,(\ref{NSIop}) indicates the modified
matter potential
\begin{eqnarray}
{\cal A} \equiv
\sqrt{2} G_F N_e \left(
\begin{array}{ccc}
1+ \epsilon_{ee} & \epsilon_{e\mu} & \epsilon_{e\tau}\\
\epsilon_{\mu e} & \epsilon_{\mu\mu} & \epsilon_{\mu\tau}\\
\epsilon_{\tau e} & \epsilon_{\tau\mu} & \epsilon_{\tau\tau}
\end{array}
\right),
\label{matter-np}
\end{eqnarray}
$\epsilon_{\alpha\beta}$ is defined by
\begin{eqnarray}
&{\ }&\hspace{-22mm}
\epsilon_{\alpha\beta}\equiv\sum_{f=e,u,d}\frac{N_f}{N_e}\epsilon_{\alpha\beta}^{f}\,,
\label{eab}
\end{eqnarray}
and $N_f~(f = e, u, d)$ is the number density of fermions $f$.
Here, we define the new NSI parameters as
$\epsilon_{\alpha\beta}^{fP}\equiv\epsilon_{\alpha\beta}^{ffP}$ and
$\epsilon_{\alpha\beta}^{f}\equiv\epsilon_{\alpha\beta}^{fL}+\epsilon_{\alpha\beta}^{fR}$
because the matter effect is sensitive only to the coherent scattering
and vector part in the interaction. As shown in
the definition of $\epsilon_{\alpha\beta}$, the neutrino oscillation
experiments conducted on Earth are only sensitive to the sum of
$\epsilon_{\alpha\beta}^{f}$.

\subsection{Solar neutrinos\label{sol_nu}}

In Refs.\,\cite{deHolanda:2010am,Gonzalez-Garcia:2013usa}
it was pointed out that a
tension occurs between the two mass squared differences
extracted from the KamLAND and solar neutrino experiments.
The mass squared difference $\Delta m^2_{21}$
($=4.7\times10^{-5} {\rm eV}^2$) extracted from the
solar neutrino data 
is $2\sigma$ smaller than that
from the KamLAND data $\Delta m^2_{21}$ ($=7.5\times10^{-5} {\rm eV}^2$).
The authors of Refs.\,\cite{Gonzalez-Garcia:2013usa}
indicated that the tension can be removed by introducing an NSI in propagation.

To discuss the effect of the NSI on the solar neutrinos, we
reduce the $3 \times 3$ Hamiltonian in the Dirac equation,
Eq.\,(\ref{eqn:sch}), to an effective $2 \times 2$ Hamiltonian to obtain
the survival probability $P(\nu_e \rightarrow \nu_e)$ because solar
neutrinos are approximately driven by one mass squared difference
$\Delta m_{21}^2$\,\cite{Gonzalez-Garcia:2013usa}.  The survival probability $P(\nu_e \rightarrow
\nu_e)$ can be written as follows:
\begin{eqnarray}
&{\ }&\hspace{-22mm}
P(\nu_e \rightarrow \nu_e) = c_{13}^4 P_{\rm eff} +  s_{13}^4\,.
\end{eqnarray}
Here, $P_{\rm eff}$ can be calculated using the effective $2 \times 2$ Hamiltonian $H^{\rm eff}$, which is written as
\begin{widetext}
\begin{eqnarray*}
H^{\rm eff}=
\frac{\Delta m^2_{21}}{4E}\left(\begin{array}{cc}
-\cos2\theta_{12} & \sin2\theta_{12}  \\
\sin2\theta_{12} & \cos2\theta_{12}
\end{array}\right) 
+
\left(\begin{array}{cc}
c^2_{13} A & 0 \\
0 & 0
\end{array}\right)  + 
 A\sum_{f=e,u,d} \frac{N_f}{N_e}
\left(\begin{array}{cc}
- \epsilon_D^f &  \epsilon_N^f \\
 \epsilon_N^{f*} &  \epsilon_D^f
\end{array}\right),
\end{eqnarray*}
where  $\epsilon^f_{D}$ and $\epsilon^f_{N}$ are linear combinations of the standard NSI parameters:
\begin{eqnarray}
\epsilon_D^f &=&c_{13}s_{13}{\rm Re}\left[ e^{i\delta_{\rm CP}}\left(s_{23}\epsilon_{e \mu}^f+c_{23}\epsilon_{e \tau}^f\right) \right]-\left(1+s_{13}^2\right)c_{23}s_{23}{\rm Re}\left[\epsilon_{\mu \tau}^f\right] \nonumber\\ 
&-&\frac{c_{13}^2}{2}\left(\epsilon_{e e}^f-\epsilon_{\mu \mu}^f\right)+\frac{s_{23}^2-s_{13}^2c_{23}^2}{2}\left(\epsilon_{\tau \tau}^f-\epsilon_{\mu \mu}^f\right) 
\label{epsilond} \\
\epsilon_N^f&=& c_{13}\left(c_{23}\epsilon_{e \mu}^f-s_{23}\epsilon_{e\tau}^f \right)+s_{13}e^{-i\delta_{\rm CP}}\left[ s_{23}^2\epsilon_{\mu\tau}^f-c_{23}^2\epsilon_{\mu\tau}^{f*} +c_{23}s_{23}\left(\epsilon_{\tau \tau}^f-\epsilon_{\mu \mu}^f\right) \right].
\label{epsilonn}
\end{eqnarray}
\end{widetext}
Ref.\,\cite{Gonzalez-Garcia:2013usa} discussed the sensitivity of solar neutrinos and KamLAND experiments to $\epsilon_D^f$ and a real $\epsilon_N^f$ for either $f = u$ or $f = d$ at a particular time.
The best fit values from the solar neutrino and KamLAND data are $(\epsilon_D^u,\epsilon_N^u)=(-0.22,-0.30)$ and $(\epsilon_D^d,\epsilon_N^d)=(-0.12,-0.16)$, whereas those
from the global analysis of the neutrino oscillation data are $(\epsilon_D^u,\epsilon_N^u)=(-0.140, -0.030)$ and $(\epsilon_D^d,\epsilon_N^d)=(-0.145, -0.036)$.
These results give us a hint regarding the existence of the NSI.
In addition to the above, Ref.\,\cite{Gonzalez-Garcia:2013usa} also discussed the possibility of 
a dark-side solution ($\Delta m^2_{21}<0$ and $\theta_{21}>45^\circ$), which requires the NSI to be used in the solar neutrino problem.
The allowed regions for the dark-side solution are disconnected from those for the standard LMA solution in 
the $(\epsilon_D^f,\epsilon_N^f)$ plane, whereas those for the dark-side solution within $3\sigma$ do not contain the standard scenario $\epsilon_D^f=\epsilon_N^f=0$.\footnote{
The COHERENT experiment has ruled out the dark-side solution\,\cite{Coloma:2017ncl}.  In Ref.\,\cite{Liao:2017uzy}, the authors discussed the constraint from the COHERENT data on the NSI.}
Furthermore, in Ref.\,\cite{Esteban:2018ppq} more general NSI
couplings to $u$ and $d$ quarks were considered as follows:
\begin{eqnarray}
\epsilon_{\alpha\beta}
=\sqrt{5}\,\left(
\cos\eta+Y_n\sin\eta
\right)\, \epsilon_{\alpha\beta}^\eta
=\sqrt{5}\,\frac{\sin(\eta+\eta_0)}{\sin(\eta_0)}\,
  \epsilon_{\alpha\beta}^\eta\,,
\end{eqnarray}
where $\epsilon_{\alpha\beta}^\eta$ is the overall
normalization of the NSI coupling,
$\eta$ is a new parameter used to interpolate $f = u$ ($\eta = \tan^{-1}(1/2) = 26.6^\circ$) and
$f=d$ ($\eta=\tan^{-1}(2)=63.4^\circ$), or
$f=p$ ($\eta=0$) and
$f=n$ ($\eta=\pi/2$),
and $\eta_0$ is
defined from the neutron-proton ratio
$Y_n\equiv\#(n)/\#(p)$ by
$\eta_0\equiv \tan^{-1}(1/Y_n)$.
It was concluded that the point with the best fit
in the global analysis is
\begin{eqnarray}
 &{\ }&\hspace{-6mm}  
  \eta=-\left.\eta_0\right|_{\text{mantle}}
  =-\left.\tan^{-1}\left(\frac{1}{Y_n}\right)\right|_{\text{mantle}}
  =- \tan^{-1}\left(\frac{1}{1.051}\right)=-43.6^\circ\,,
  \label{eta0}
\end{eqnarray}
i.e., as far as the terrestrial experiments are
concerned, the scenario without the NSI provides the best fit.
In fact, if the condition (\ref{eta0}) is satisfied,
then the NSI effect disappears in the terrestrial experiments,
including T2HKK and DUNE, as discussed below.
The condition (\ref{eta0}), however,
is a certain type of fine tuning, and for
a generic value such as $\eta=\tan^{-1}(1/2)=26.6^\circ$
($f=u$) or $\eta=\tan^{-1}(2)=63.4^\circ$ (f=d),
the NSI effect does not disappear in the terrestrial experiments.
Thus, in the following, we adopt $\eta=\tan^{-1}(1/2) = 26.6^\circ$
or $\eta=\tan^{-1}(2)=63.4^\circ$ as typical reference values
for $\eta$.

\subsection{T2HKK and DUNE}
\label{t2hk-dune}

The T2HKK experiment\,\cite{Abe:2016ero}
is a proposal for the future extension
of the T2K experiment\,\cite{Abe:2014tzr}.\footnote{
The possibility of a second detector in Korea for the T2K
experiment was previously discussed \cite{Kim:2000kias,Hagiwara:2004iq,Ishitsuka:2005qi,Hagiwara:2005pe,Hagiwara:2006vn,
Kajita:2006bt,Barger:2007jq,Kimura:2007mu,Ribeiro:2007jq,
Huber:2007em,Hagiwara:2009bb,Oki:2010uc,Hagiwara:2011kw,Hagiwara:2012mg,Dufour:2012zr,Hagiwara:2016qtb}.}
 Under this proposal, a water \cnv\,\,detector with a fiducial mass of 187 kt
is placed not only in Kamioka (at a baseline length $L$ of 295 km) but also
in Korea (at $L$ of $\simeq$1,100 km); in addition, the power of the beam
at J-PARC in Tokai Village is upgraded to 1.3 MW.
In a similar manner as the off-axis design ($2.5^\circ$) used in the T2K experiment, it is assumed that
T2HKK uses an off-axis beam at a 1.5$^\circ$
angle between the directions
of the decaying charged pions and neutrinos,
and the neutrino energy spectrum has a peak at approximately
0.8 GeV.

By contrast, DUNE\,\cite{Acciarri:2015uup} is another long-baseline experiment planned in the USA.  Its baseline length
and peak energy are $L$=1,300 km and $E\sim3$ GeV, respectively.
It will be driven by a 1.2 MW proton beam, and is designed to
accommodate future beam power upgrades to 2.4 MW.
It is expected that a liquid argon detector with a fiducial mass of 40 kt will provide
information for a wide range of the neutrino energy.

The matter effect appears in the neutrino oscillation probability,
typically in the form of
$G_FN_eL/\sqrt{2}$ = $[\rho/(2.6g/cm^3)][L/(4000\mbox{\rm km})]$.
The baseline length of T2HK ($L$ = 295 km) is too short for the matter
effect,
and thus T2HK has poor sensitivity to NSIs in the neutrino propagation.
The baseline lengths of T2HKK and DUNE are
comparable to the typical length, which is estimated
based on the matter effect, and thus T2HKK and DUNE are expected to be
sensitive to NSIs in the neutrino propagation.

We used GLoBES \cite{Huber:2004ka,Huber:2007ji} and MonteCUBES \cite{Blennow:2009pk} software to simulate all the above experiments described above. The run time of both DUNE and T2HKK are considered to be 10
years. For T2HKK the ratio of neutrino and antineutrino running is 1:3, whereas for DUNE it is 1:1. Our results are consistent with Refs. \cite{Abe:2016ero,Acciarri:2015uup}. We can calculate our 
sensitivity in terms of $\chi^2$ in the following way: 
\begin{eqnarray}
 &{\ }&\hspace{-6mm}
  \chi^2_{{\rm stat}} = 2 \sum_i \left\{ 
 \tilde{N}^{{\rm test}}_i - N^{{\rm true}}_i 
 - N^{{\rm true}}_i \log\left(
 \frac{\tilde{N}^{{\rm test}}_i}
 {N^{{\rm true}}_i}\right) \right\}.
 \label{chi2stat}
 \end{eqnarray}
 The index $i$ corresponds to the number of energy bins,
 and $\tilde{N}^{{\rm test}}_i$ indicates the test events
 obtained through the original test events $N^{{\rm test}}_i$
 by a scale factor used to incorporate the effect of systematic errors
 in the following manner:
 \begin{eqnarray}
  \tilde{N}^{{\rm test}}_i \equiv 
 \left(1+ \sum_k c_i^k \xi_k  \right)
 N^{{\rm test}}_i 
 \end{eqnarray}
 where $c_i^k$ is the $1 \sigma$ systematic error corresponding to the
 pull variable $\xi_k$, and the index $k$ indicates the number of pull
 variables. The final $\chi^2$ is obtained by varying $\xi_k$ from $-3$ to $+3$, corresponding to their $3 \sigma$ ranges and minimizing the combination of the
 statistical ($\chi^2_{{\rm stat}}$)
 and systematic ($\sum_k \xi_k^2$)
 contributions over $\xi_k$ as well as the oscillation parameters:
 \begin{eqnarray}
  \chi^2 = \displaystyle\min_{\xi_k,~\mbox{\rm\scriptsize osc.~param}}\left(\chi^2_{\rm stat}  + \sum_k \xi_k^2 \right).
 \label{chi2}
 \end{eqnarray}
 For T2HKK we took an
overall systematic error of 3.2\% (3.6\%) for the appearance (disappearance) channel in neutrino
mode and 3.9\% (3.6\%) for the appearance (disappearance) channel in antineutrino mode. The systematic error is the same for both the signal and background.
The systematic error for DUNE is 2\% (10\%)
for the appearance channel and 5\% (15\%) for the disappearance channel corresponding to the signal
(the background). The systematic errors in neutrino and antineutrino mode are the same
for DUNE. 

\section{Correspondence between the long baseline
and solar neutrino experiments\label{sec3}}

\begin{figure*}[t]
\hspace{-3.2mm}
\includegraphics[scale=0.23]{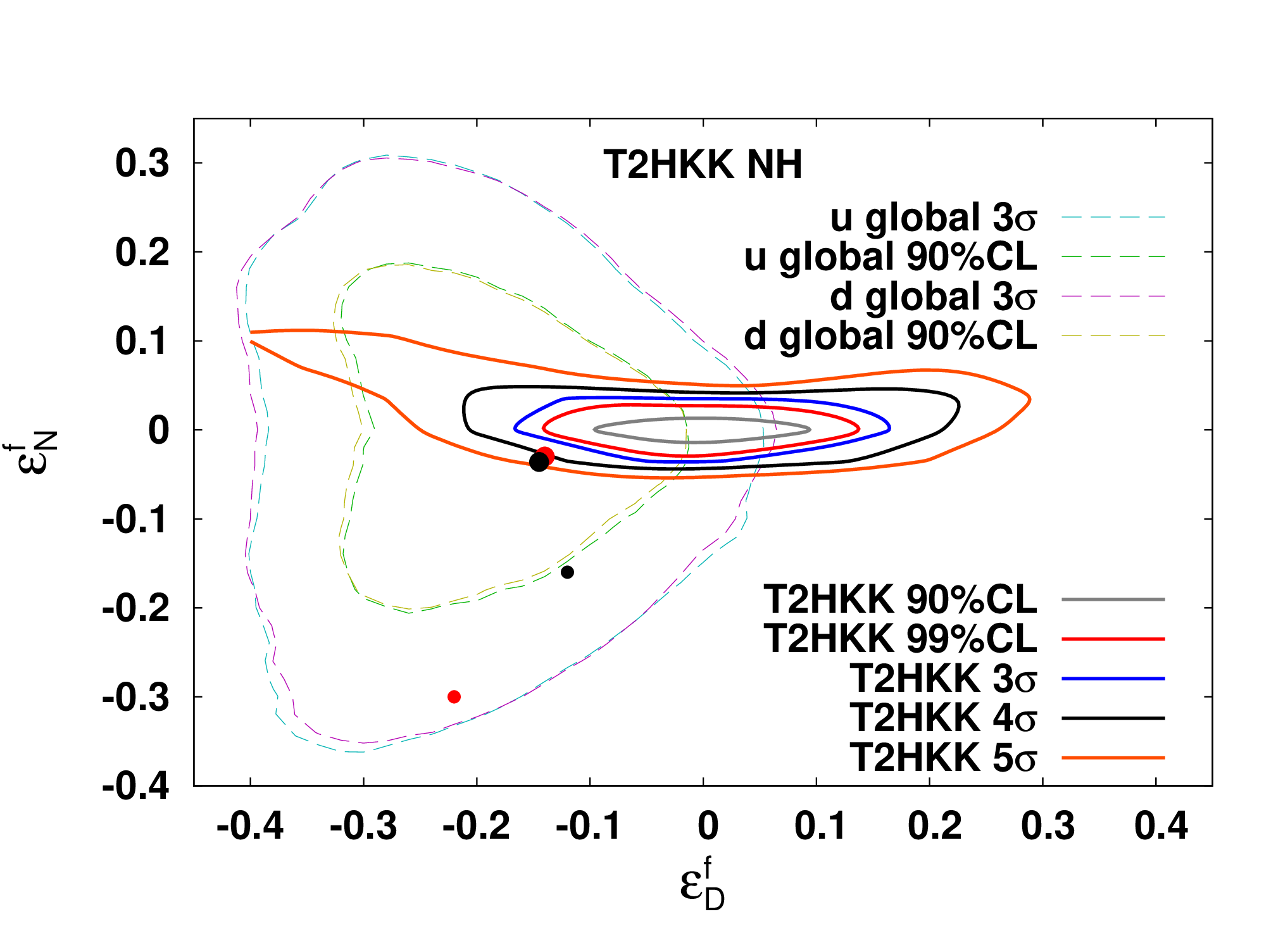}
\hspace{-3.2mm}
\includegraphics[scale=0.23]{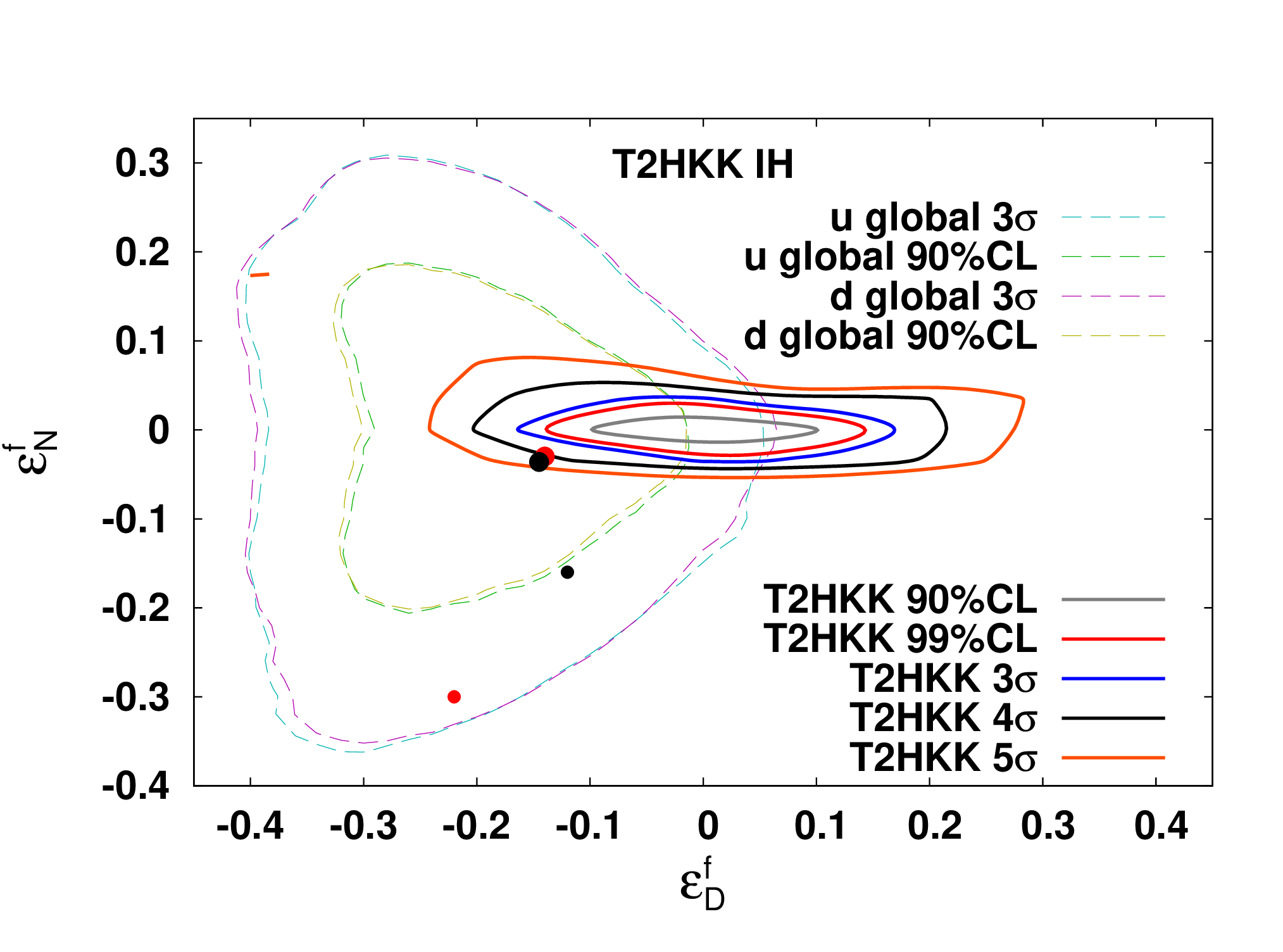}
\hspace{-3.2mm}
\includegraphics[scale=0.23]{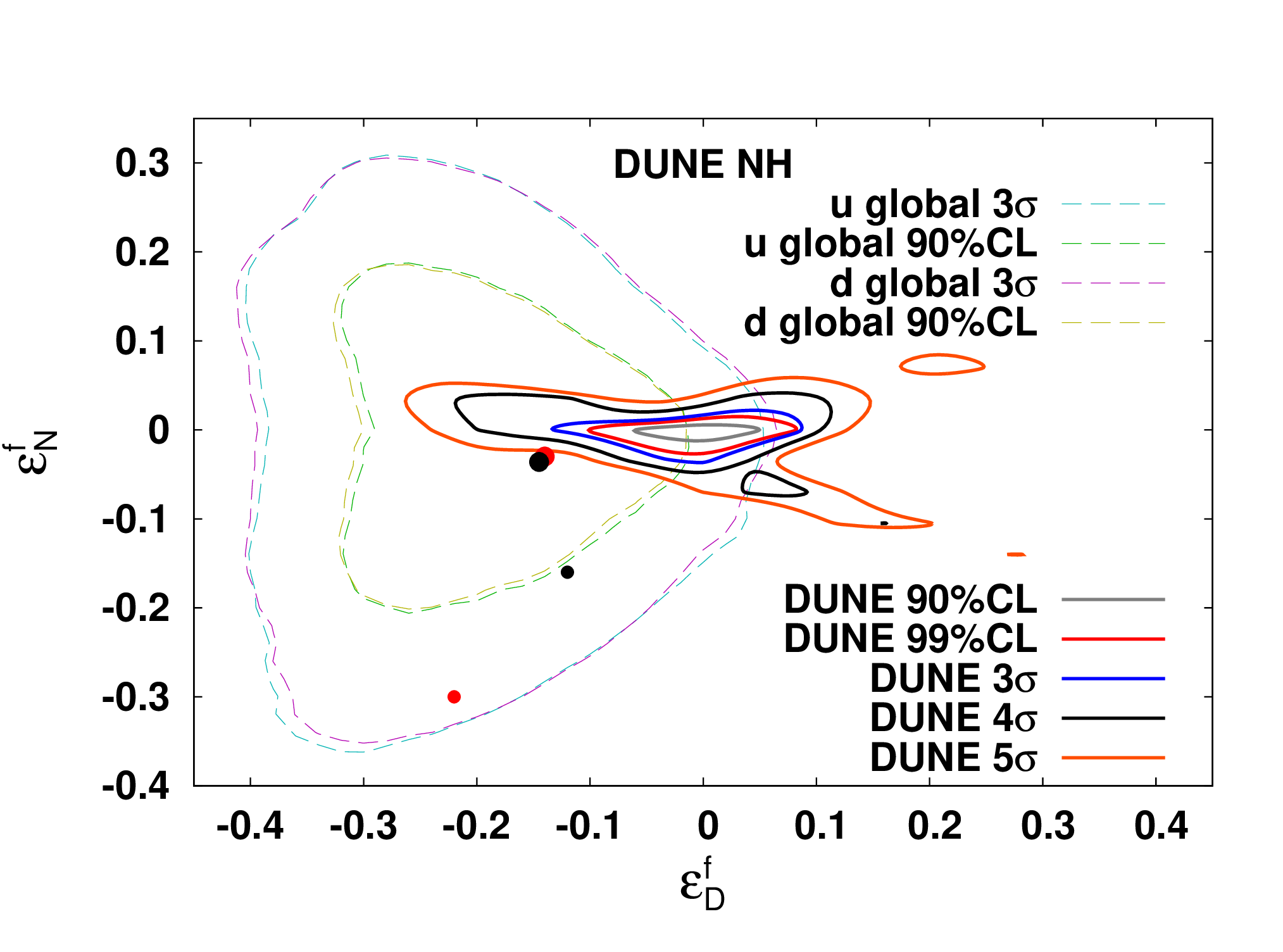}
\hspace{-3.2mm}
\includegraphics[scale=0.23]{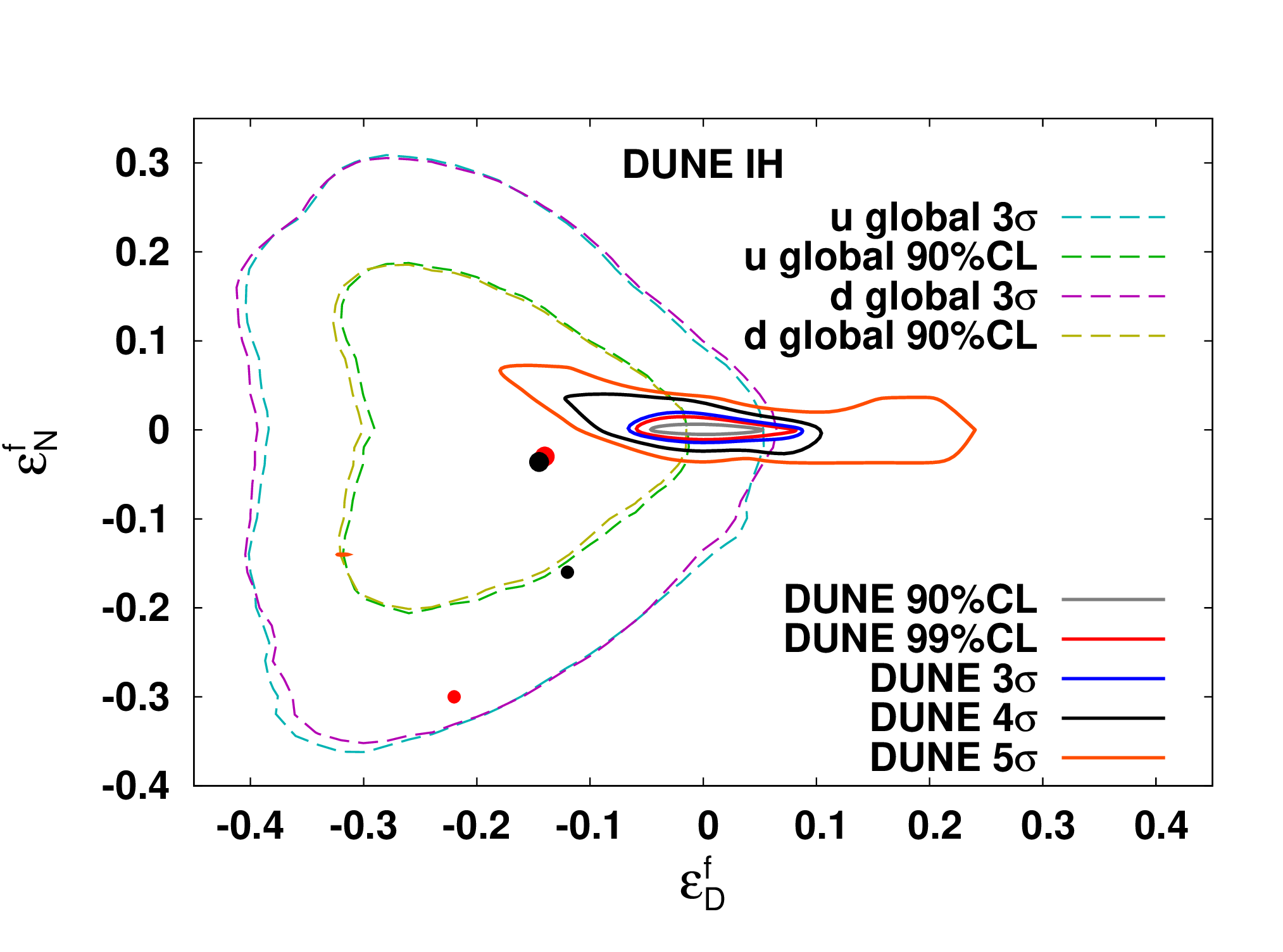}
\caption{Excluded regions in ($\epsilon_D$, $\epsilon_N$) plane 
for T2HKK and DUNE for $\dcp = -90^\circ$ and $\theta_{23}=45^\circ$
(outside of the curves).  The allowed regions at
90\%CL and 3$\sigma$ suggested by the global
analysis\,\cite{Gonzalez-Garcia:2013usa} are
also shown in the dashed curves (inside of the curves).  The large
(small) red and black circles indicate
the best fit points for
$f=u$ and $f=d$ from the global (solar + KamLAND)
analysis\,\cite{Gonzalez-Garcia:2013usa}, respectively.
} 
\label{fig:fig1}
\end{figure*}

Our strategy applied in this study is to provide the
excluded region within the ($\epsilon_D^f$, $\epsilon_N^f$) plane
by marginalizing all the parameters.
This means that, given a set of parameters
($\epsilon_D^f$, $\epsilon_N^f$),
we have to minimize $\chi^2$ by varying all
parameters that satisfy the relations
(\ref{epsilond}) and (\ref{epsilonn}).
For this purpose,
let us discuss the relation of the NSI parameters $\epsilon_{\alpha\beta}$
to $\epsilon_D^f$ and $\epsilon_N^f$.
The first thing to note is that the relation between
$\epsilon_{\alpha\beta}$ and ($\epsilon_D^f$, $\epsilon_N^f$)
is a many-to-one mapping,
and thus we have to choose the independent and dependent
variables from the two relations (\ref{epsilond}) and
(\ref{epsilonn}).  In the following, we will
treat $\epsilon_{ee}^{f}$, 
$|\epsilon_{e\tau}^{f}|$ and $\epsilon_{\tau\tau}^{f}$
as dependent parameters and regard all others, namely,
$\theta_{23}$, $\dcp$, 
$|\epsilon_{e\mu}^{f}|$, arg($\epsilon_{e\mu}^{f}$),
arg($\epsilon_{e\tau}^{f}$), $|\epsilon_{\mu\tau}^{f}|$,
and arg($\epsilon_{\mu\tau}^{f}$) as independent variables.\footnote{
The errors of the standard oscillation parameters
$\theta_{12}$, $\Delta m_{21}^2$, $\Delta m_{32}^2$, and $\theta_{13}$
have little impact on our analysis, and thus we will
fix these parameters throughout this study.}
The second point to mention is that
the constraint on $\epsilon_{\mu\mu}$ is so
strong\,\cite{Davidson:2003ha,Biggio:2009nt}
that $|\epsilon_{\mu\mu}|$ is much smaller than
the errors of $\epsilon_{ee}^{f}$
and $\epsilon_{\tau\tau}^{f}$.  Thus we
can assume that $\epsilon_{\mu\mu}=0$ has
a good approximation in Eqs. (\ref{epsilond}) and
(\ref{epsilonn}).
Because the analysis in Ref.\,\cite{Gonzalez-Garcia:2013usa}
was applied for the real
$\epsilon_N^f$, Eq.\,(\ref{epsilonn}) implies that
the real part of the right-hand side of Eq.\,(\ref{epsilonn})
equals $\epsilon_N^f$, whereas the imaginary part of
the right-hand side of Eq.\,(\ref{epsilonn}) disappears.
From Eq.\,(\ref{epsilonn}), we can therefore express
$|\epsilon_{e\tau}^{f}|$ and $\epsilon_{\tau\tau}^{f}$
in terms of the other parameters as follows:
\begin{eqnarray}
|\epsilon_{e\tau}^f| &=& 
\frac{1}
{c_{13}c_{23}\sin(\phi_{13}+\delta_{\rm CP})}\,
\left(-F\sin\delta_{\rm CP}+G\cos\delta_{\rm CP}\right)
\nonumber\\
\hspace{-90pt}
\epsilon_{\tau \tau}^f
&=& \frac{2}
{s_{13}\sin2\theta_{23}\,\sin(\phi_{13}+\delta_{\rm CP})}\,
\left(F\sin\phi_{13}+G\cos\phi_{13}\right)
\nonumber
\end{eqnarray}
where $\phi_{jk}$, $F$, and $G$ are defined in the following manner:
\begin{eqnarray}
\phi_{12} &\equiv& \mbox{\rm arg}(\epsilon_{e\mu}^f),~~
\phi_{13} \equiv \mbox{\rm arg}(\epsilon_{e\tau}^f),~~
\phi_{23} \equiv \mbox{\rm arg}(\epsilon_{\mu\tau}^f),
\nonumber\\
F&\equiv& \epsilon_{N}^f
-c_{13}c_{23}\,|\epsilon_{e\mu}^f|\,\cos\phi_{12} \\ \nonumber
&-& s_{13}|\epsilon_{\mu\tau}^{f}|
\left\{s_{23}^2\cos(\phi_{23}-\delta_{\rm CP})
-c_{23}^2\cos(\phi_{23}+\delta_{\rm CP})\right\}
\label{epsnr}\\ 
G &\equiv&  -c_{13}c_{23}\,|\epsilon_{e\mu}^f|\,\sin\phi_{12} \\ \nonumber
&-&s_{13}|\epsilon_{\mu\tau}^{f}|
\left\{s_{23}^2\sin(\phi_{23}-\delta_{\rm CP})
+c_{23}^2\sin(\phi_{23}+\delta_{\rm CP})\right\}\,.
\label{epsni}
\end{eqnarray}
After we obtain $|\epsilon_{e\tau}^{f}|$ and $\epsilon_{\tau\tau}^{f}$,
we obtain $\epsilon_{ee}^{f}$ from Eq.\,(\ref{epsilond}):
\begin{eqnarray} \nonumber
\epsilon_{e e}^f
 &=&\frac{2}{c_{13}^2}
\left\{
\frac{s_{23}}{2}\sin2\theta_{13}|\epsilon_{e \mu}^f|
\cos(\delta_{\rm CP}+\phi_{12}) \right. \\ \nonumber
&+& \left. \frac{c_{23}}{2}\sin2\theta_{13}|\epsilon_{e \tau}^f|
\cos(\delta_{\rm CP}+\phi_{13})\right.
\nonumber \\
&-& \left.\left(1+s_{13}^2\right)c_{23}s_{23}|\epsilon_{\mu \tau}^f|
\cos(\phi_{23}) \right. \\ \nonumber
&-& \left. \epsilon_D^f
+\frac{s_{23}^2-s_{13}^2c_{23}^2}{2}\epsilon_{\tau \tau}^f
\right\}
\label{epsilond1}
\end{eqnarray}
\begin{table*}
\begin{center}
\begin{tabular}{|c|c|c|c|c|c|c|c|c|c|c|}
\hline
$\epsilon_{ee}$&$|\epsilon_{e\tau}|$& $\epsilon_{\tau\tau}$&$\dcp$&$\theta_{23}$&arg($\epsilon_{e\tau}$)&$|\epsilon_{\mu\tau}|$&arg($\epsilon_{\mu\tau}$)&$|\epsilon_{e\mu}|$&arg($\epsilon_{e\mu})$&$\chi^2$  \\          
\hline
 0.846   &   0.123   &  -0.021     &   -90   &    47    &   0     &       0         &        0         &     0       &   0       &   25.46 \\
1.128  &     0.108  &   0.511     &      -90  &    45   &    30   &      0.15        &     90       &     0      &    0        &  17.54      \\
0.917  &    0.146     &  0.114      &     -90   &    47   &   0        &    0        &         0        &    0.03   &   30      &   24.61 \\

\hline
\end{tabular}
\end{center}
\caption{The values of the oscillation parameters and $\chi^2$ for ($\epsilon_D$, $\epsilon_N$) =(-0.14, -0.03) corresponding to T2HKK and NH.
}
\label{tab1} 
\end{table*}

\begin{figure*}
\begin{center}
\hspace{-3.2mm}
\includegraphics[scale=1.04]{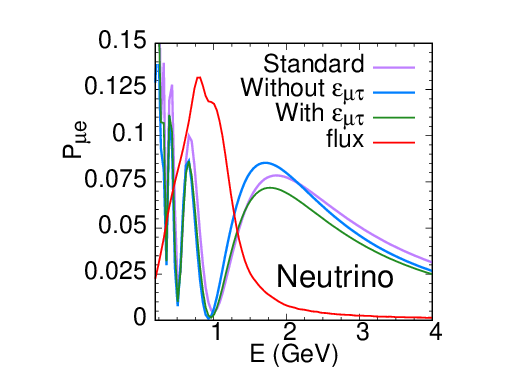}
\hspace{-20.2mm}
\includegraphics[scale=1.04]{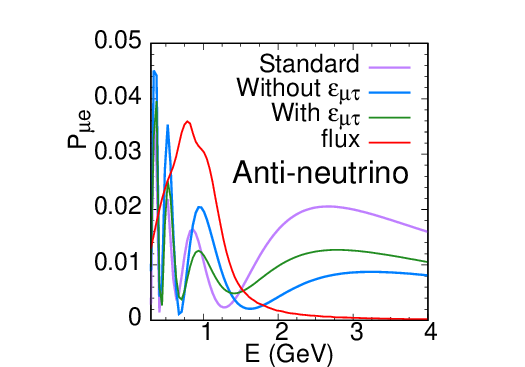}
\caption{The appearance probability 
$P(\nu_\mu\to\nu_e)$ with and without
the NSI parameters.
The flux at the detector in Korea is
also shown.}
\label{fig:fig2}
\end{center}
\end{figure*}

\begin{figure*}
\hspace{-3.2mm}
\includegraphics[scale=0.23]{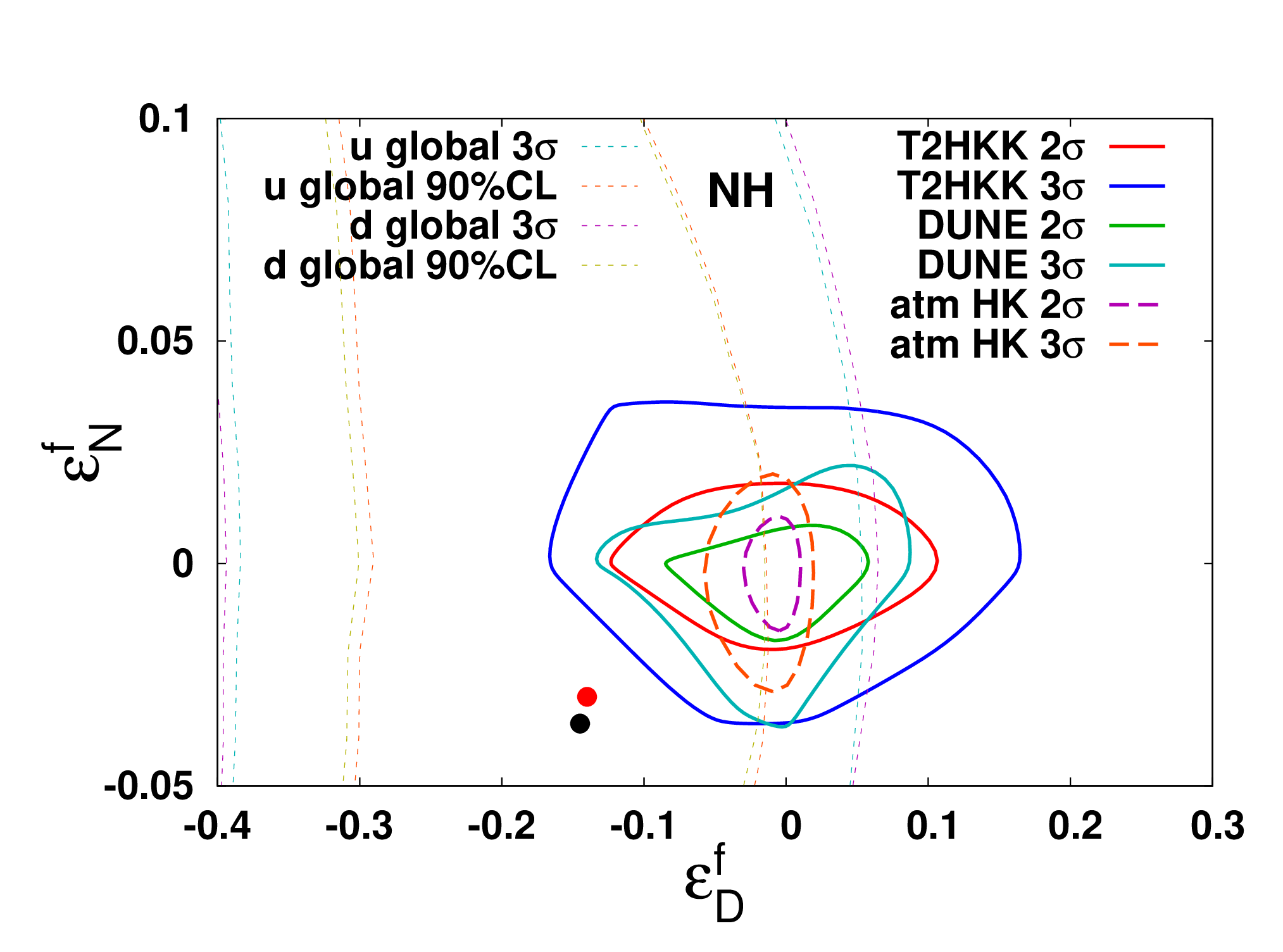}
\hspace{-3.2mm}
\includegraphics[scale=0.23]{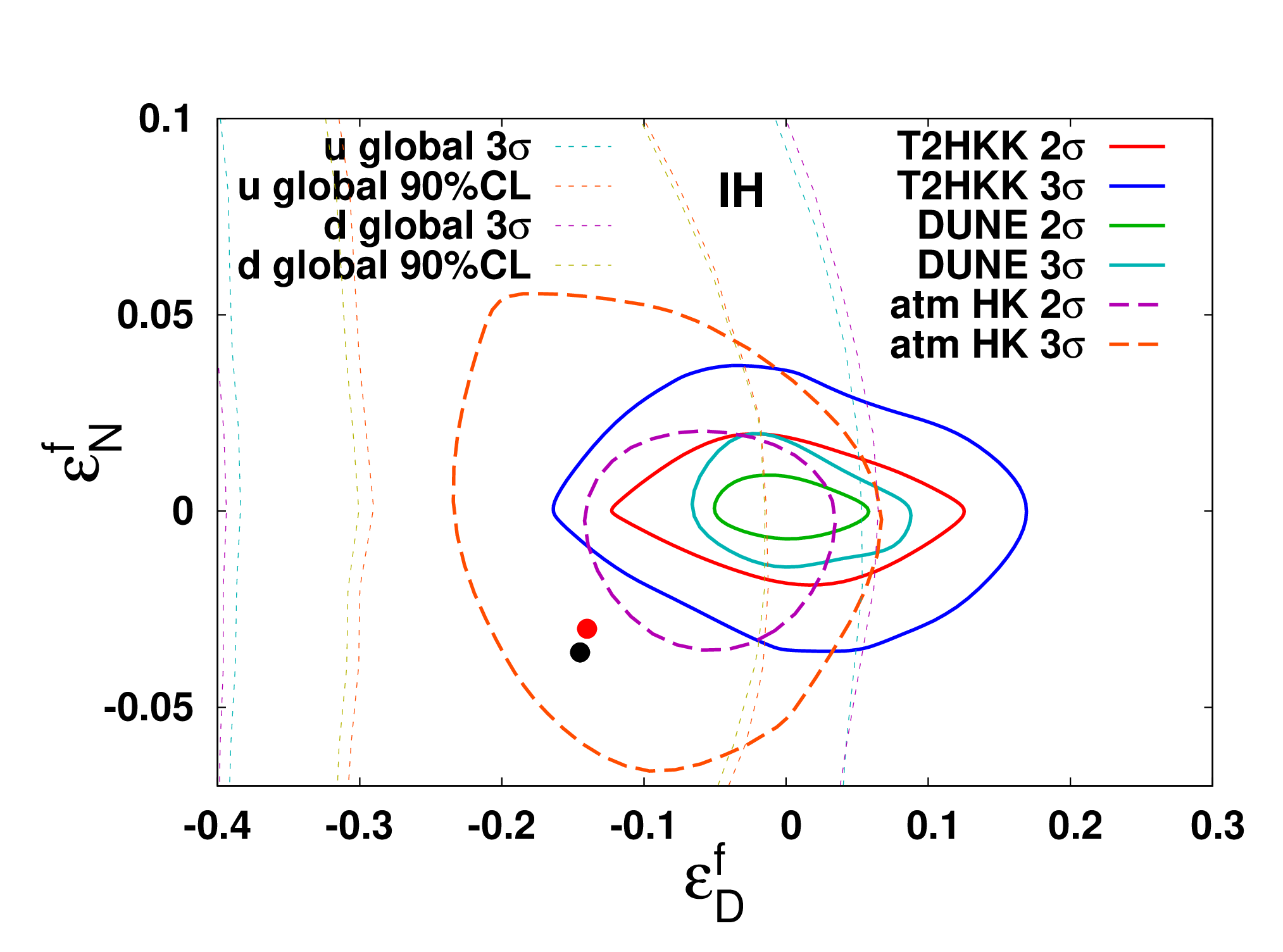}
\caption{Comparison of the excluded regions by T2HKK, DUNE, and
HK atmospheric neutrino observations\,\cite{Fukasawa:2016nwn,Fukasawa:2017thesis}
in the ($\epsilon_D$, $\epsilon_N$) plane.
Others are the same as in Fig.\,\ref{fig:fig1}.
}
\label{fig:fig3}
\end{figure*}

\begin{figure*}
\includegraphics[scale=0.237]{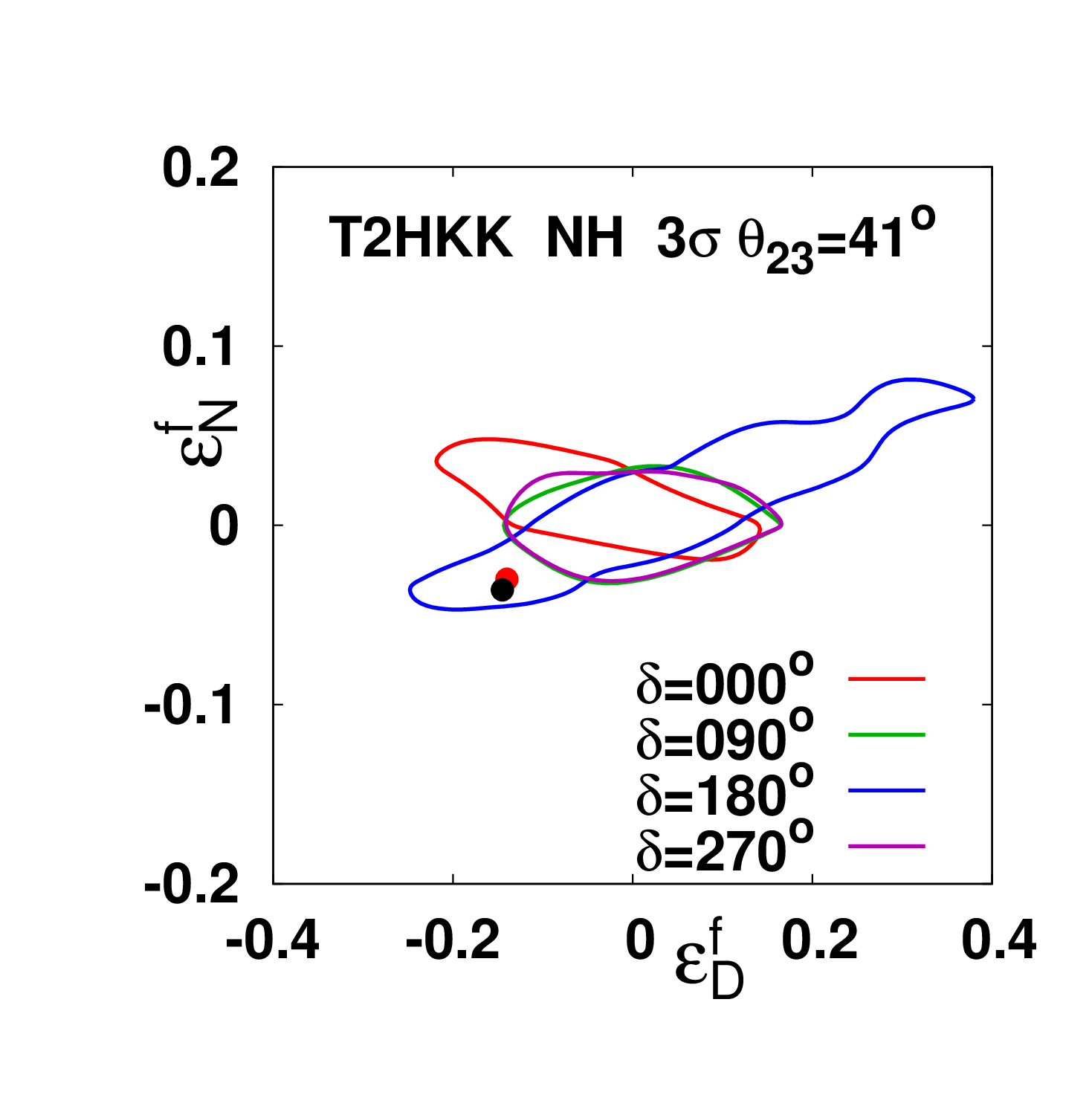}
\hspace{-8.6mm}
\includegraphics[scale=0.237]{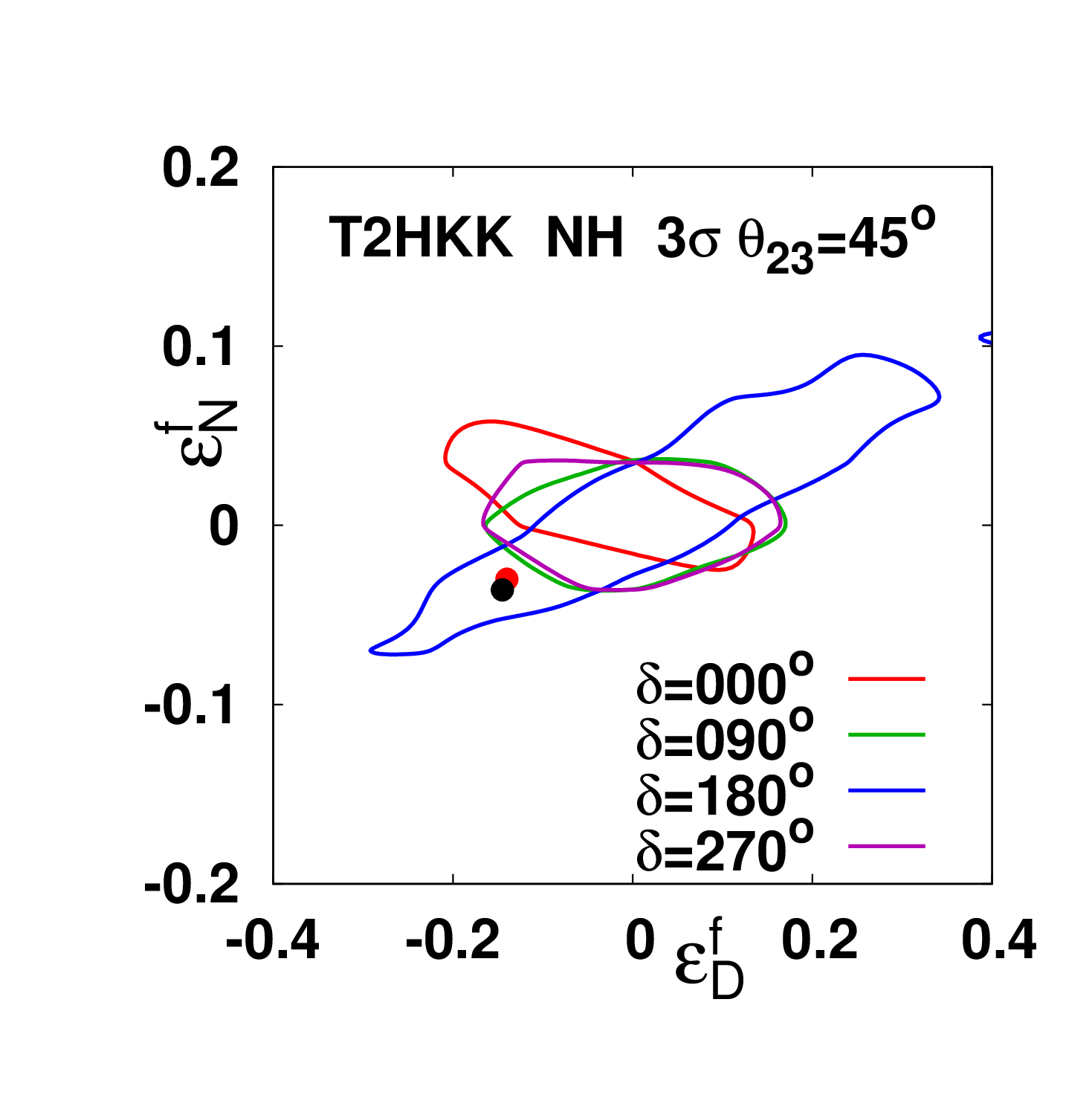}
\hspace{-8.6mm}
\includegraphics[scale=0.237]{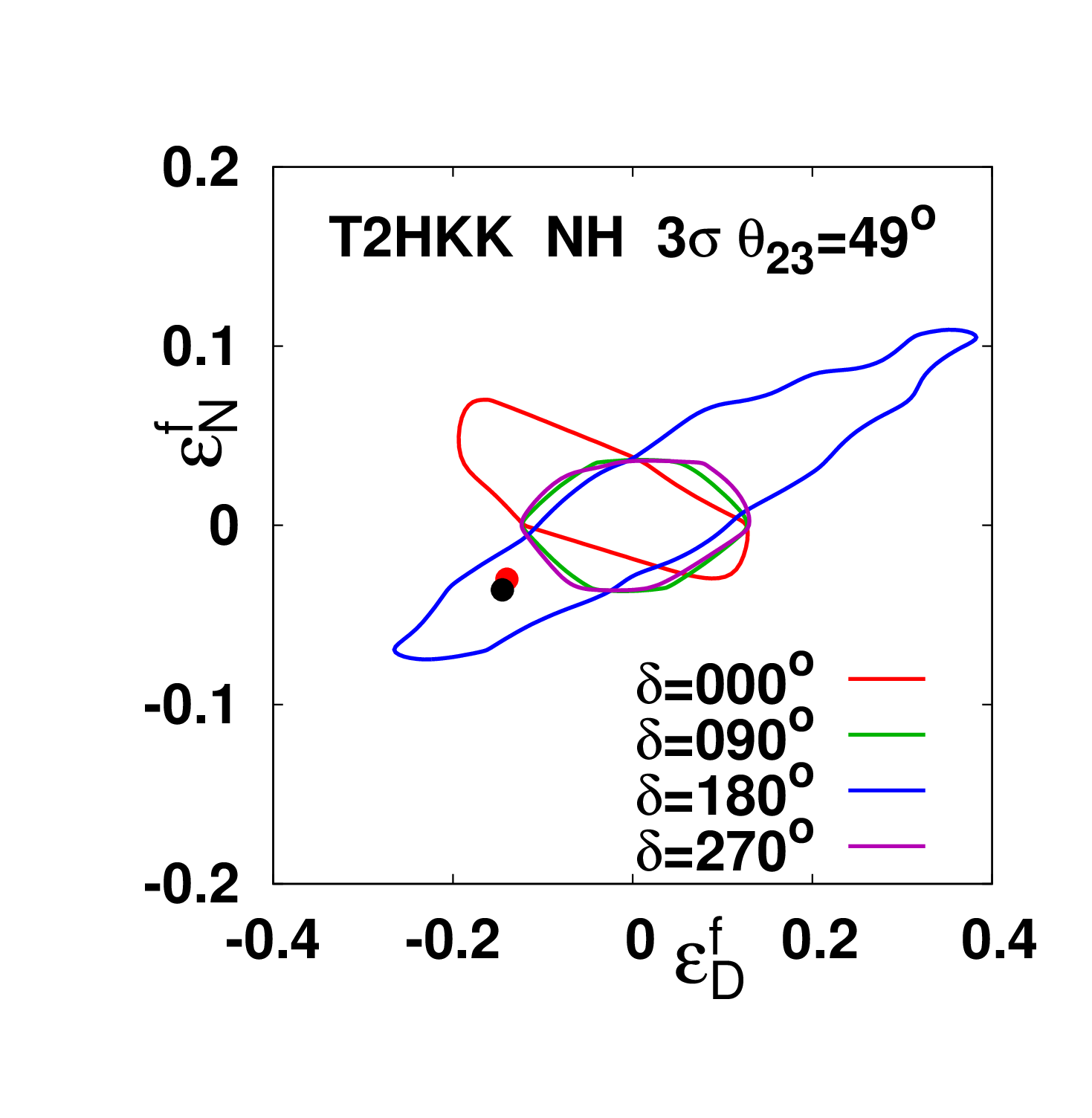}\\
\vspace{-8.2mm}
\includegraphics[scale=0.237]{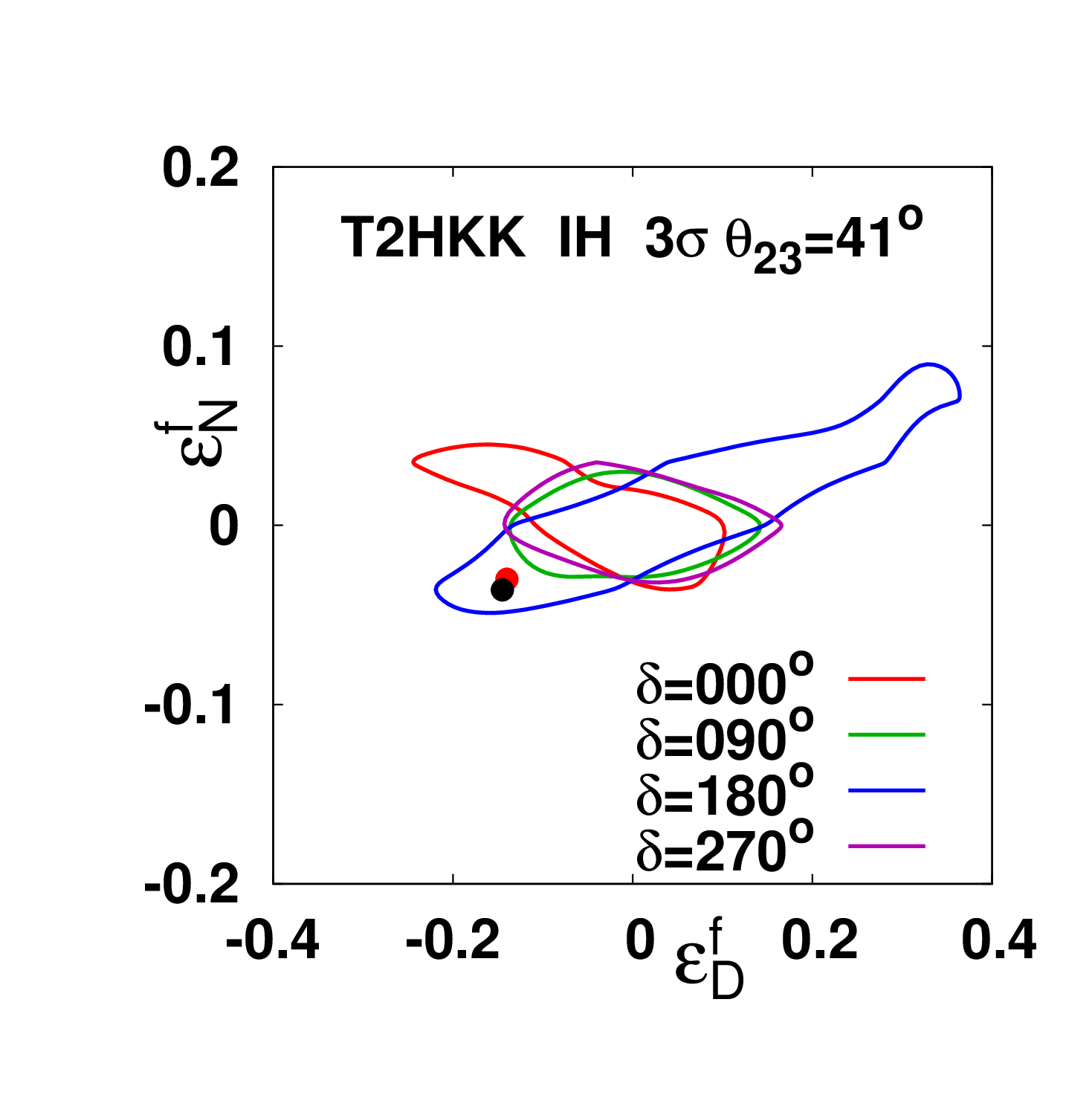}
\hspace{-8.2mm}
\includegraphics[scale=0.237]{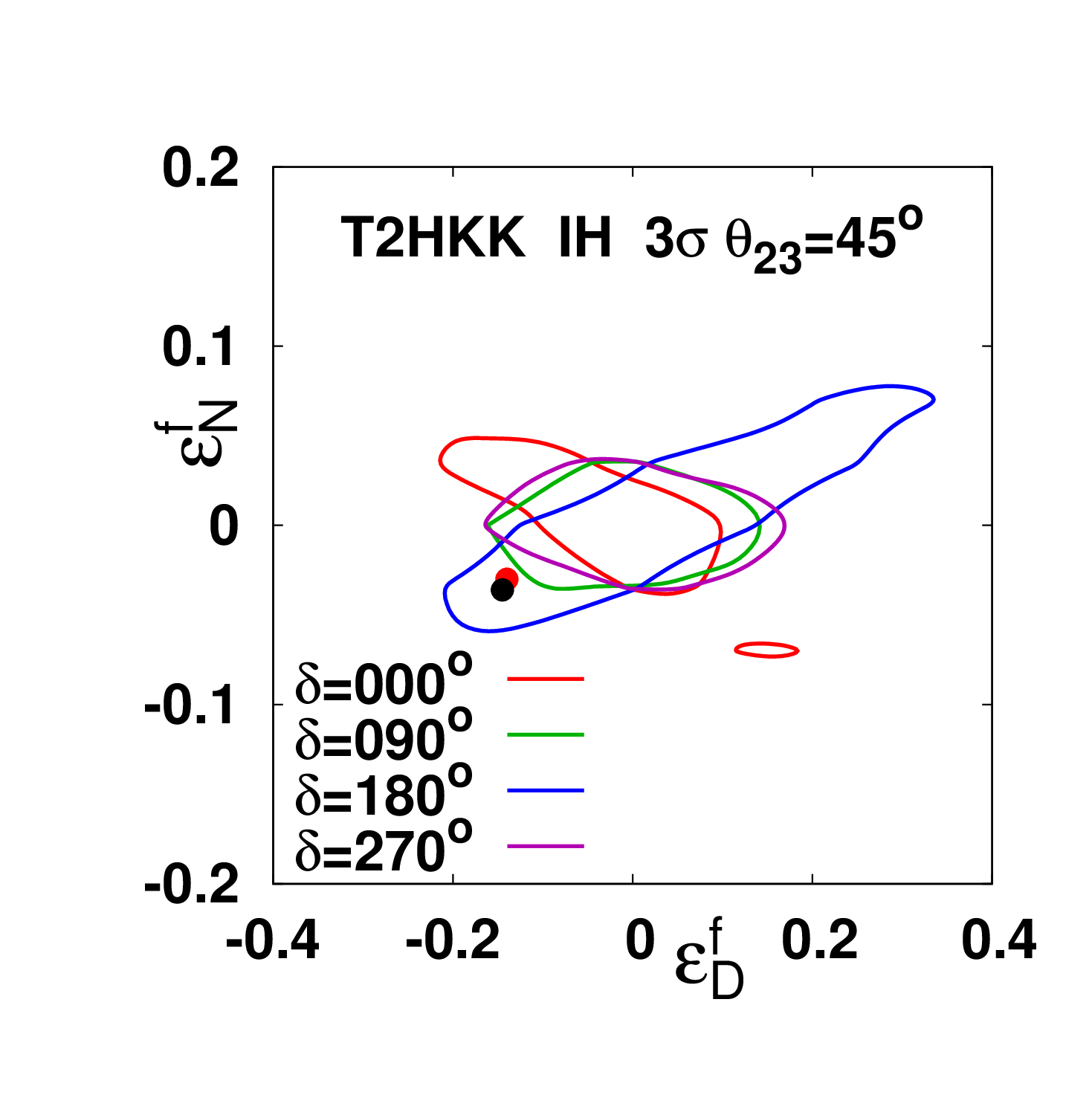}
\hspace{-8.2mm}
\includegraphics[scale=0.237]{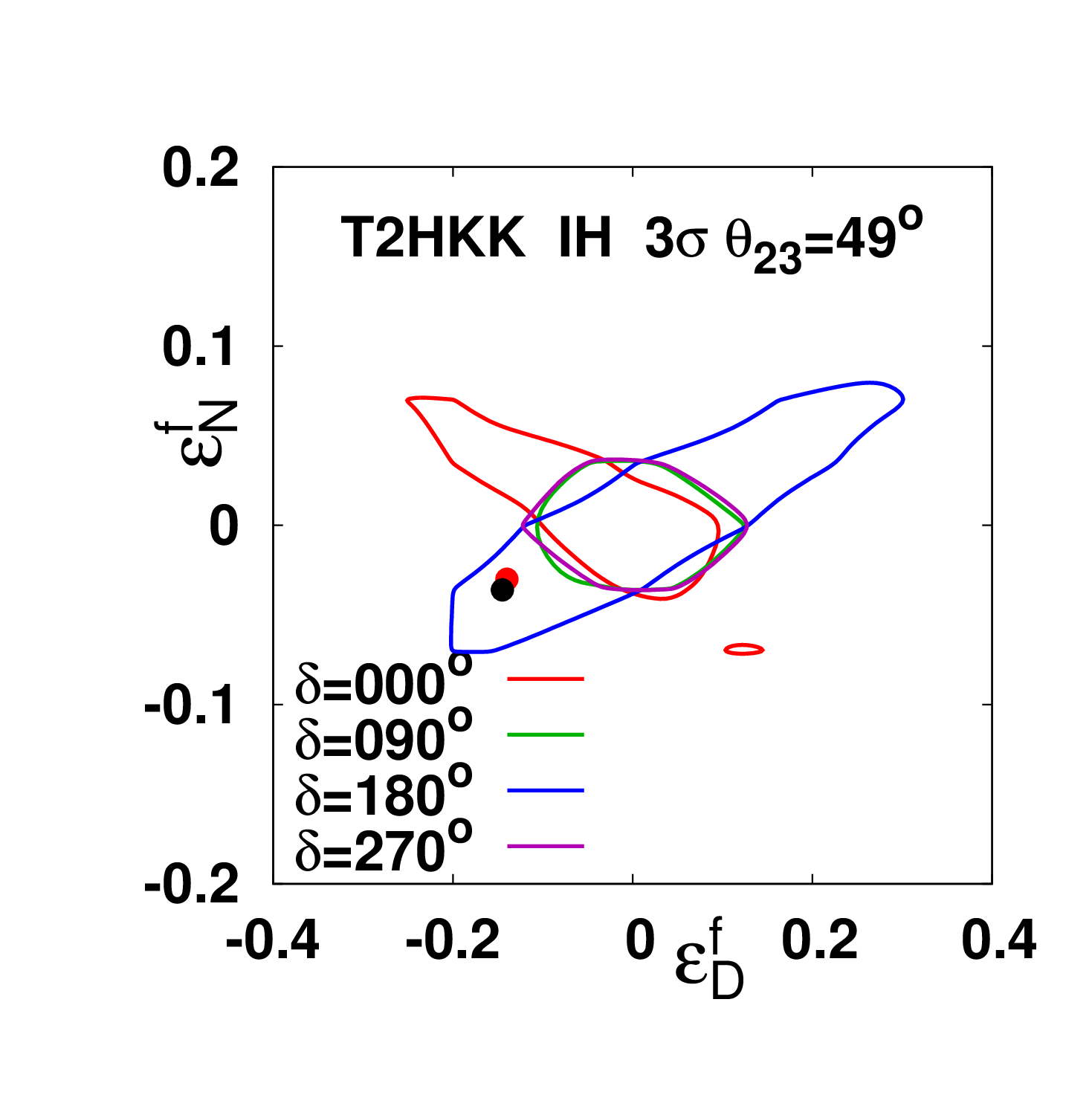}\\
\vspace{-8.2mm}
\includegraphics[scale=0.237]{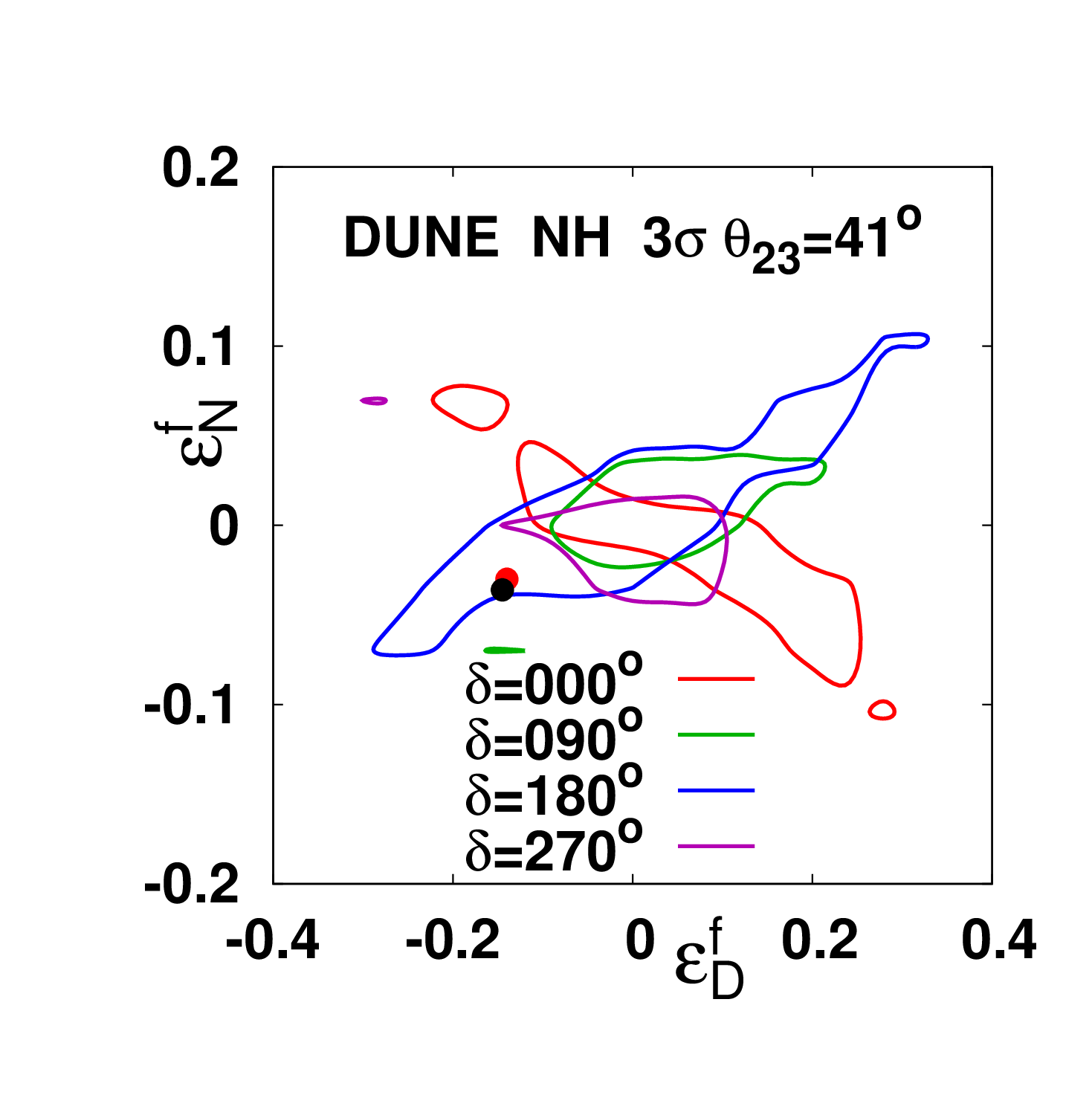}
\hspace{-8.2mm}
\includegraphics[scale=0.237]{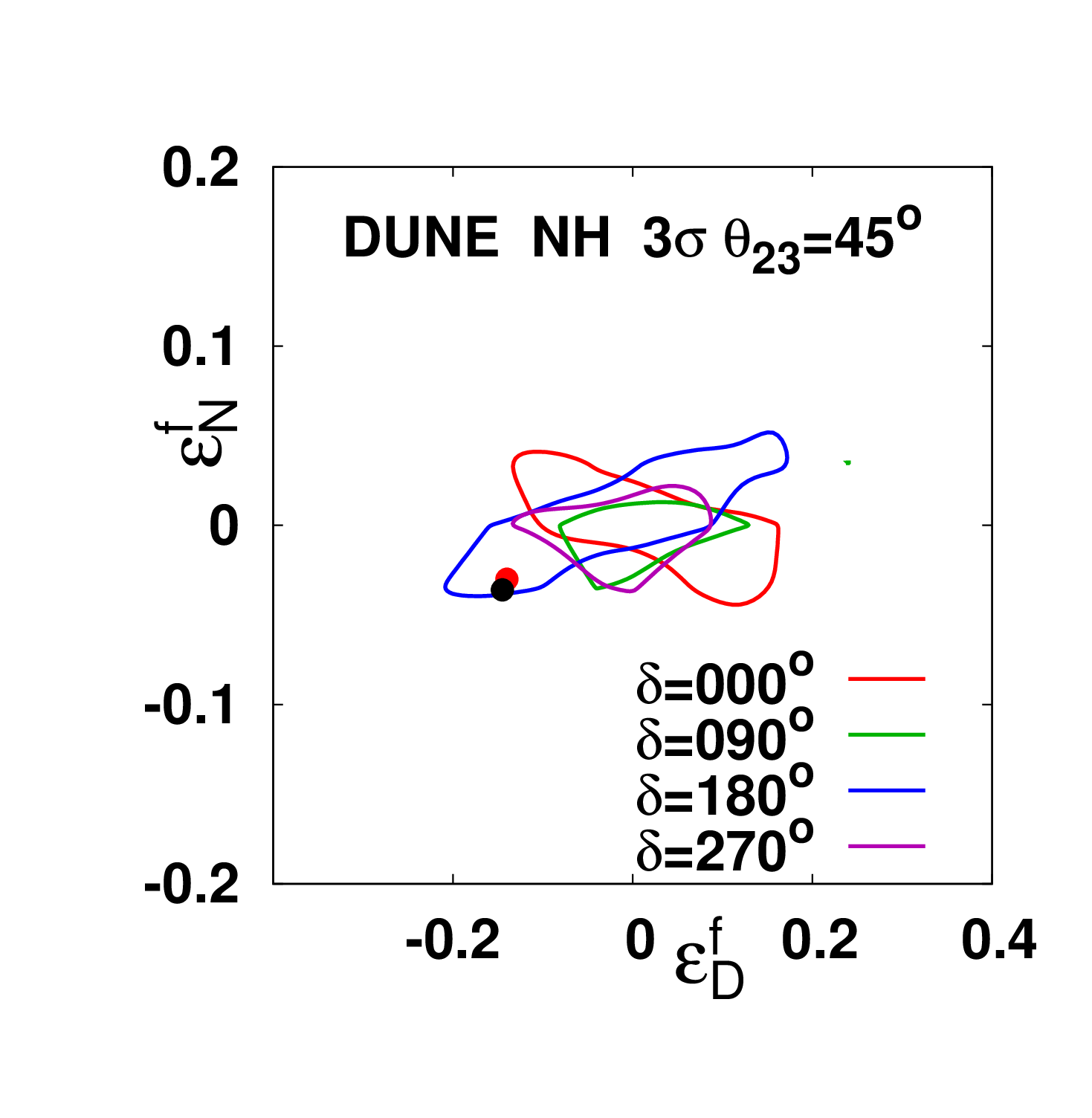}
\hspace{-8.2mm}
\includegraphics[scale=0.237]{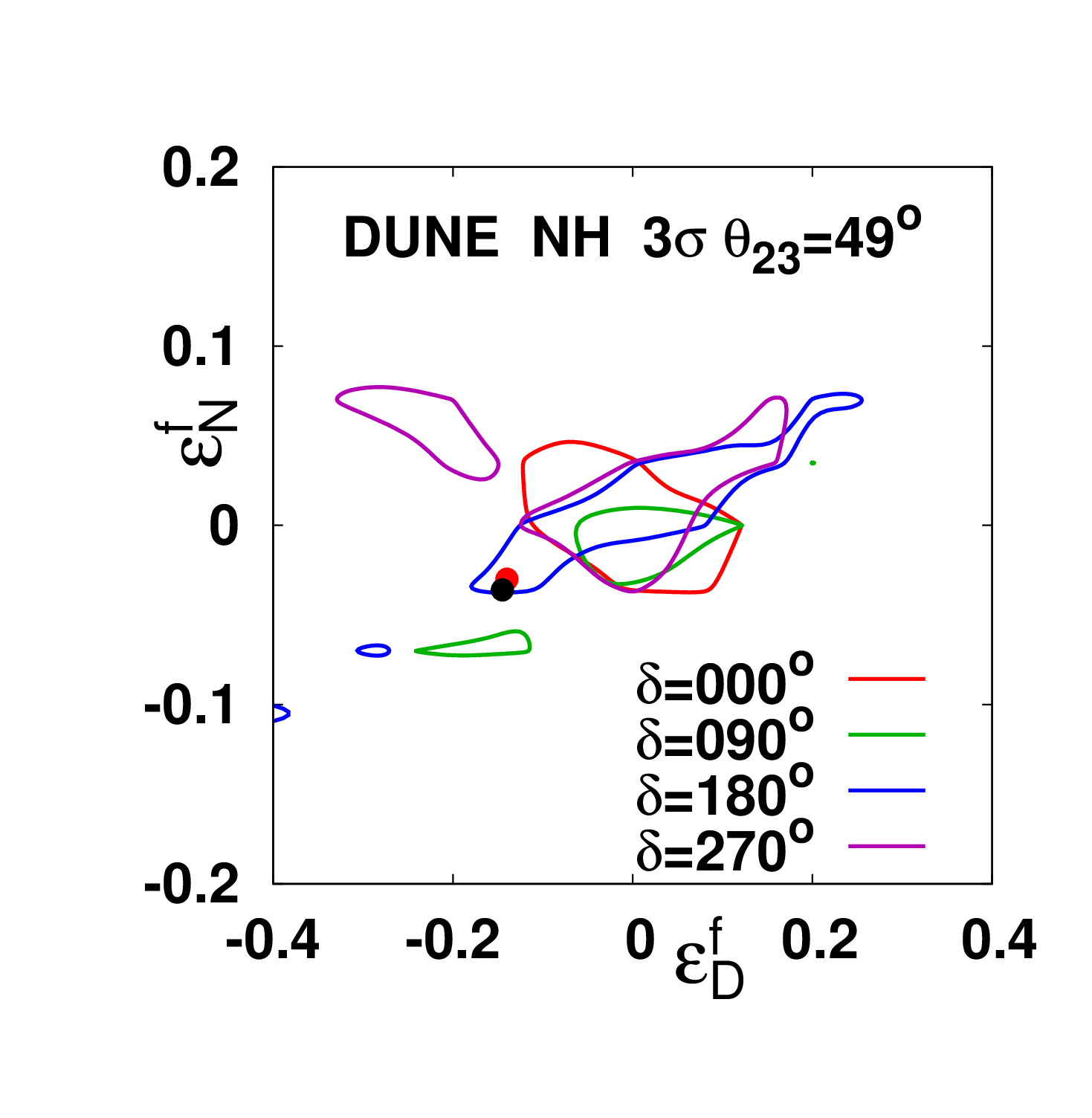}\\
\vspace{-8.2mm}
\includegraphics[scale=0.237]{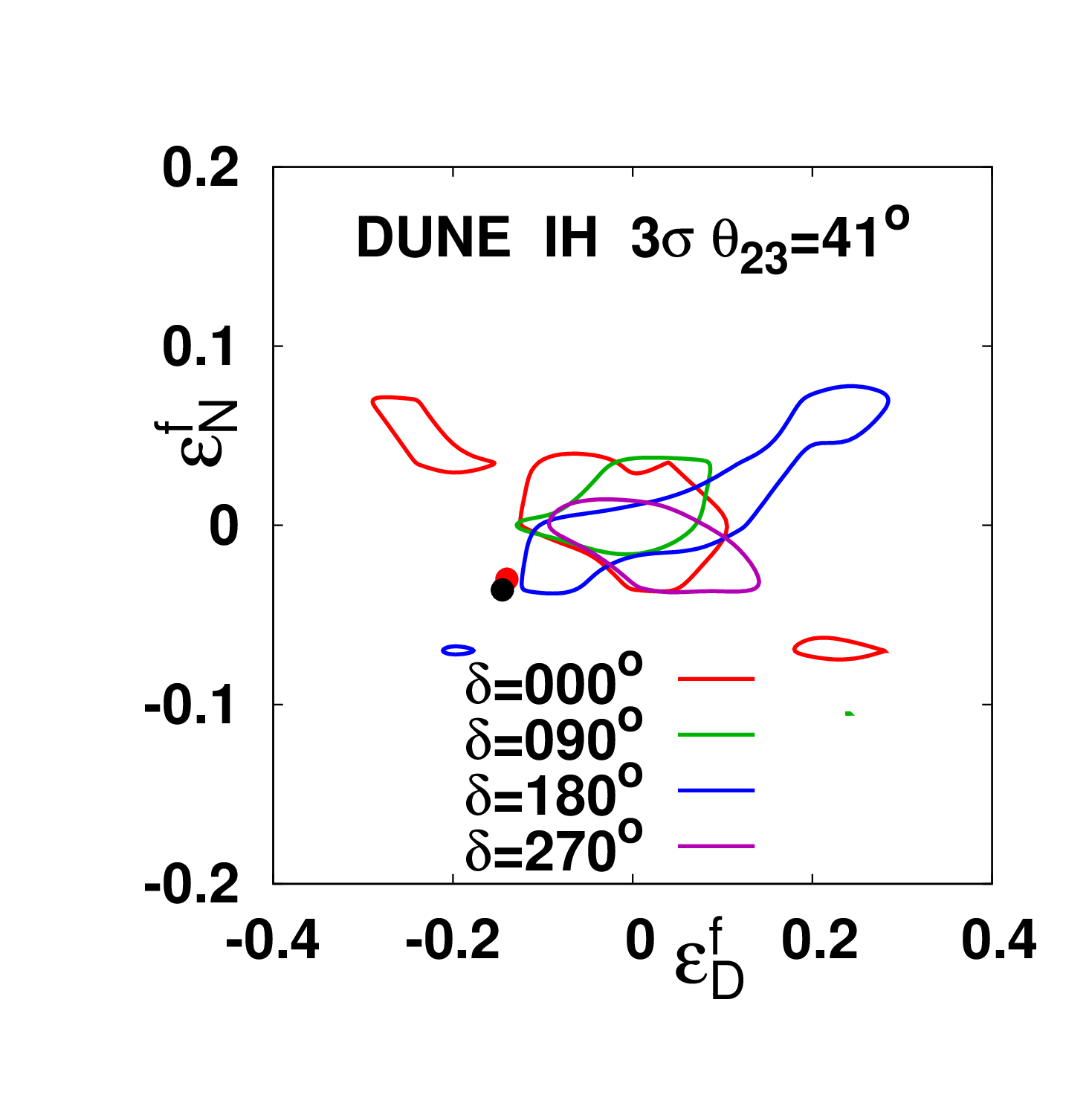}
\hspace{-8.2mm}
\includegraphics[scale=0.237]{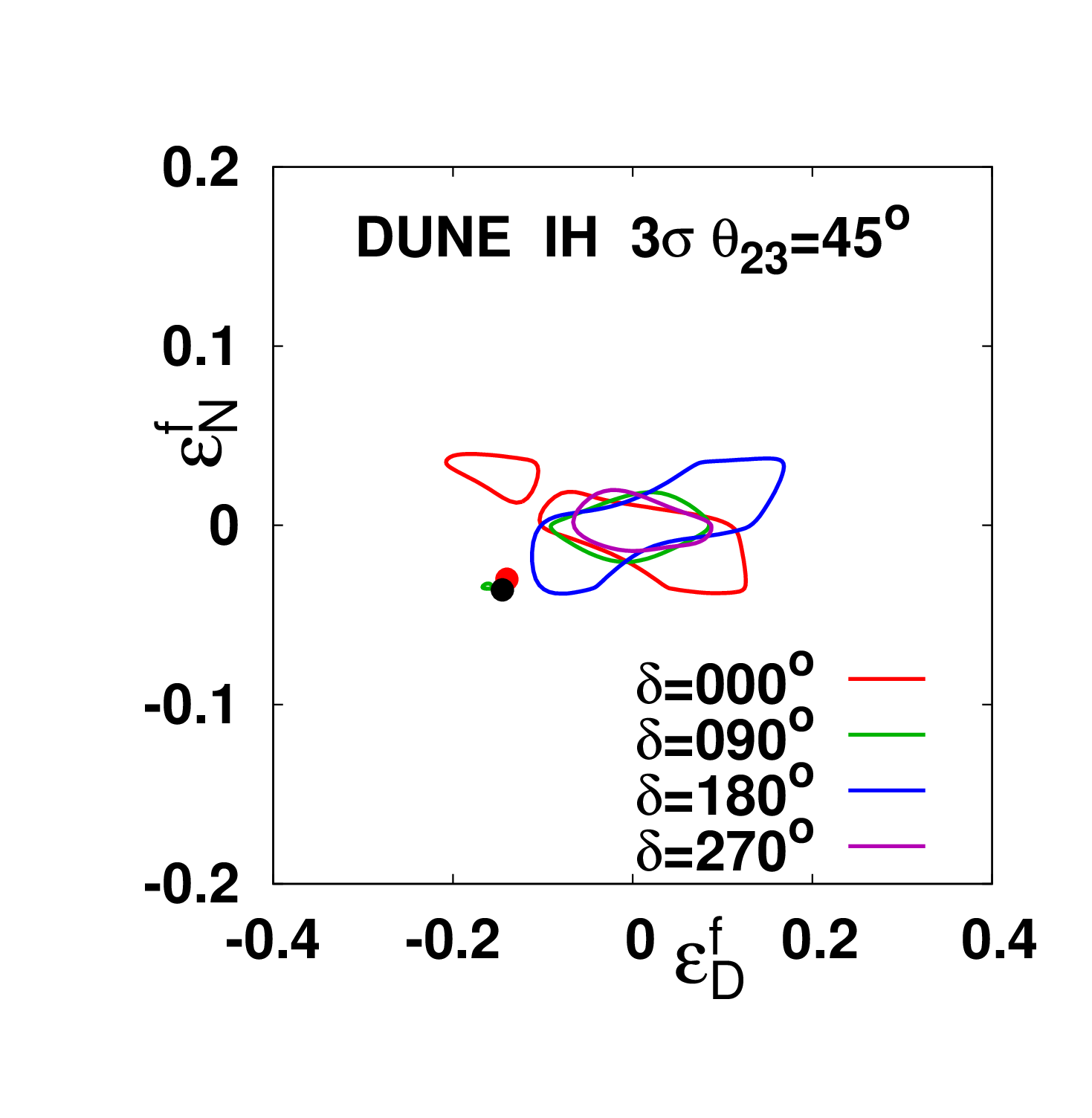}
\hspace{-8.2mm}
\includegraphics[scale=0.237]{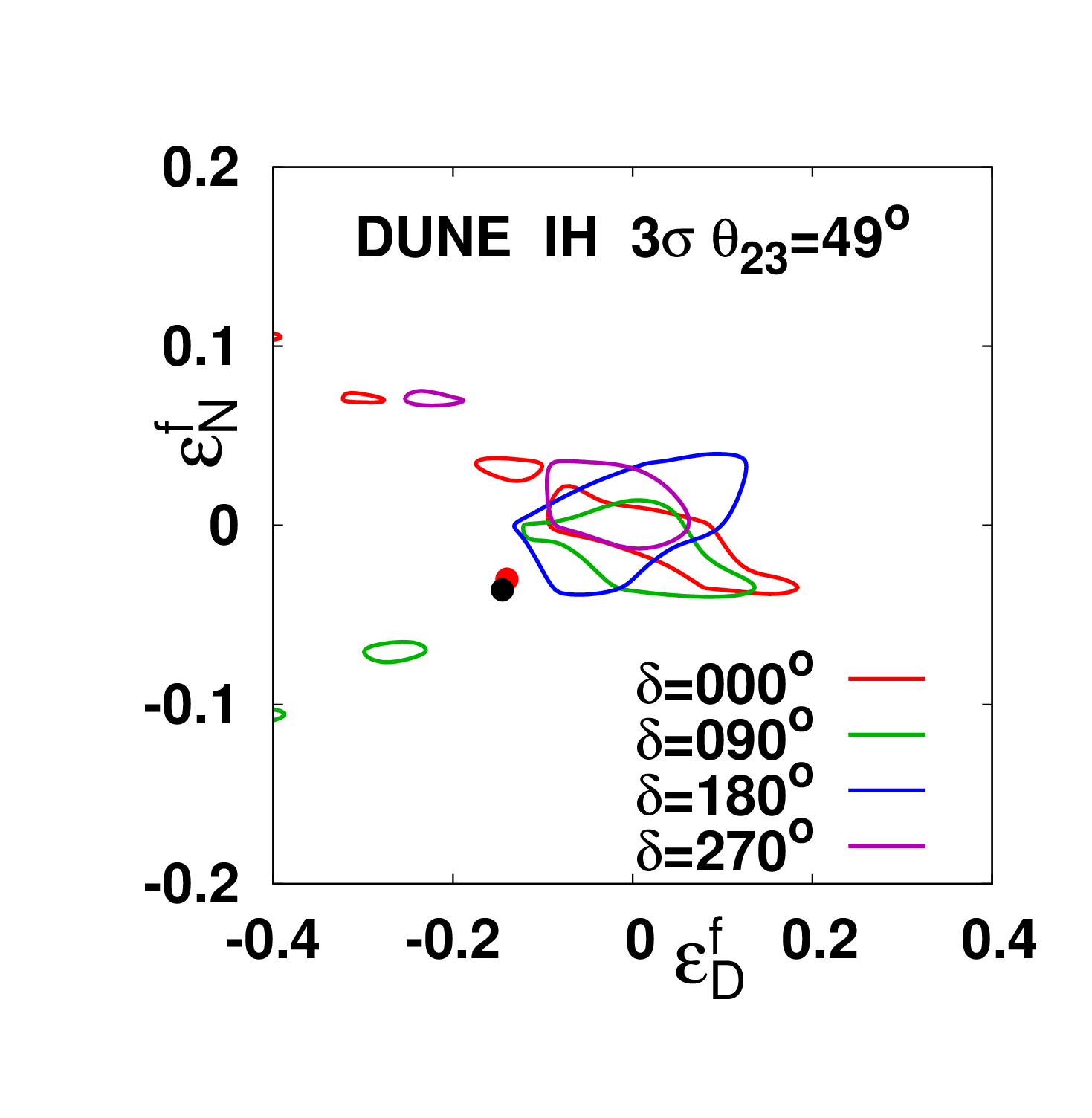}
\caption{Dependence of the $3 \sigma$ excluded region
on the true values of $\dcp$ and $\theta_{23}$ for T2HKK and DUNE.
The red and black circles indicate
the best fit points for
$f=u$ and $f=d$ from the global
analysis\,\cite{Gonzalez-Garcia:2013usa}.}
\label{fig:fig4}
\end{figure*}

\begin{table}
\begin{center}
\begin{tabular}{|c|c|c|c|}
\hline
$\dcp$ & $\theta_{23}$ & $\chi^2$ (T2HKK) & $\chi^2$ (DUNE)  \\          
\hline
 -180   &   41   & 3.45054   &  5.37943 \\ 
   &     45  &  2.42115  & 3.74085 \\
  &    49     & 4.12396  & 6.59227 \\
 \hline
 \hline
 -90   &   41   & 18.6853   &  13.0887 \\ 
   &     45  &  17.5457  &  31.8923 \\
  &    49     & 21.0747 & 21.2725 \\
 \hline
 \hline
 0   &   41   & 58.7701  &  26.9004 \\ 
   &     45  & 55.3685  &  40.5282 \\
  &    49     & 53.0394 & 34.5516 \\
 \hline
 \hline
 90   &   41   & 18.8283  &  21.0033  \\ 
   &     45  & 11.3261  &  35.3203\\
  &    49     & 23.2802 & 9.26665   \\
 \hline
\end{tabular}
\end{center}
\caption{The values of $\chi^2$ for ($\epsilon_D$, $\epsilon_N$) =(-0.14, -0.03) for T2HKK and DUNE in NH.
}
\label{tab2} 
\end{table}
Unlike in the case of the solar neutrino analysis with the NSI,
in an analysis of the oscillations in the Earth,
the ratio of electrons to nucleons
$Y_e = \#(p)/(\#(p)+\#(n))$
is approximately 1/2.
Thus, we have
\begin{eqnarray}
\epsilon_{\alpha\beta} = 3\,\epsilon_{\alpha\beta}^f
\label{f2wof}
\end{eqnarray}
for either choice of $f=u$ or $f=d$,
as can be seen from Eq.\,(\ref{eab}).
In the following, we adopt the
parameter $\epsilon_{\alpha\beta}$
which is related to $\epsilon_{\alpha\beta}^f$,
using Eq.\,(\ref{f2wof}).
The constraints on $\epsilon_{\alpha\beta}^f$
for $f=u$ and $f=d$ are given in Ref.\,\cite{Biggio:2009nt}
for terrestrial experiments, and at 90\% CL
we have
\begin{eqnarray}
&{\ }&\hspace{-20mm}
|\epsilon_{e\mu}^f| < 0.05,~~
|\epsilon_{\mu\tau}^f| < 0.05
\nonumber
\end{eqnarray}
for both $f=u$ and $f=d$.
This leads to the following prior on the moduli
of $\epsilon_{e\mu}=3\epsilon_{e\mu}^f$ and 
$\epsilon_{\mu\tau}=3\epsilon_{\mu\tau}^f$:
\begin{eqnarray}
\chi^2_{\mbox{\rm\scriptsize prior}}
=2.7\left(\frac{|\epsilon_{e\mu}|}{0.15}\right)^2
+2.7\left(\frac{|\epsilon_{\mu\tau}|}{0.15}\right)^2\,.
\nonumber
\end{eqnarray}
It is understood that this prior should be
included in $\chi^2$, i.e.,
\begin{eqnarray}
 \chi^2 = \displaystyle\min_{\xi_k,~\mbox{\rm\scriptsize osc.~param}}\left(\chi^2_{\rm stat}  + \sum_k \xi_k^2 + \chi^2_{\mbox{\rm\scriptsize prior}}
\right)
\label{chi2v2}
\end{eqnarray}

In our analysis we found that the effects
of $|\epsilon_{e\mu}^{f}|$ and arg($\epsilon_{e\mu}^{f}$)
are small compared to those of $|\epsilon_{\mu\tau}|$ and arg($\epsilon_{\mu\tau}$). 
In the following analysis, we therefore fix
the value of $|\epsilon_{e\mu}^{f}|$ to zero, which we discuss in detail in the next section.

\section{Results}
\label{sec4}

Assuming that nature is described by standard
oscillation scheme and that
the mass hierarchy is known,
we can obtain $\chi^2$ at each point
in the ($\epsilon_D^f$, $\epsilon_N^f$) plane
for T2HKK and DUNE.
The excluded regions at 90\%CL, 99\%CL, 3$\sigma$,
4$\sigma$, and 5$\sigma$
are shown in Fig.\,\ref{fig:fig1}.
The true oscillation parameters are
$\sin^22\theta_{12}=0.84$,
$\Delta m^2_{21}=7.8\times 10^{-5}$eV$^2$,
$\Delta m^2_{31}=2.5\times 10^{-3}$eV$^2$,
$\theta_{23}=45^\circ$,
$\sin^22\theta_{13}=0.09$, and
$\dcp=-90^\circ$.
For comparison, the allowed regions at 90\%CL
and 3$\sigma$ suggested by the global analysis
in Ref.\,\cite{Gonzalez-Garcia:2013usa} are
also depicted. 
The large
(small) red and black circles indicate
the best fit points for $f=u$ and $f=d$ from the global (solar + KamLAND)
analysis, respectively.
The left column is for a normal hierarchy ($\Delta m^2_{31} > 0$: NH) and the right column is for an inverted hierarchy ($\Delta m^2_{31} < 0$: IH), whereas the first row is for T2HKK and
the second row is for DUNE.

In general, the sensitivity of DUNE
is better than that of T2HKK.
We can see that both experiments
will exclude some of the regions
suggested by the global analysis,
although it is difficult for
both experiments to exclude
the region near the origin
(the standard scenario).
The best-fit point
of the combined analysis 
of the solar neutrino and KamLAND data by 
Ref.\,\cite{Gonzalez-Garcia:2013usa} can be excluded at
more than $10\sigma$, whereas the best fit point
of the global analysis in Ref.\,\cite{Gonzalez-Garcia:2013usa}
can be excluded at
$3\sigma$, by both T2HKK and DUNE.
For T2HKK the sensitivity is same for both NH and IH, whereas for DUNE the sensitivity in IH is slightly better than NH.
It is remarkable that the
excluded region is relatively horizontal,
i.e., the constraint is stronger in
the direction of $\epsilon_N^f$
compared to the one of $\epsilon_D^f$.
This is because the appearance probability
$P(\nu_\mu\to\nu_e)$ is sensitive to
$|\epsilon_{e\tau}|\sim |\epsilon_N^f|$
whereas $\epsilon_D^f\sim \epsilon_{ee}$
changes the magnitude of the matter effect,
and the accelerator-based long baseline
experiments with energy $E_\nu\sim$ of a few GeV
and baseline lengths $L\sim$(1000km) are
not very sensitive to the matter effect.

Now let us discuss the effects of $\epsilon_{e\mu}$ and $\epsilon_{\mu\tau}$. To understand these, we calculate  $\chi^2$ for the NSI parameter set
 ($\epsilon_D$, $\epsilon_N$)
 =(-0.14, -0.03), which is the best fit
 point of the global analysis for $f=u$ in 
 Ref.\,\cite{Gonzalez-Garcia:2013usa}, for three cases: (i) $|\epsilon_{e\mu}| = 0$ and $|\epsilon_{\mu\tau}| = 0$, (ii) $|\epsilon_{e\mu}| = 0$ and $|\epsilon_{\mu\tau}| \neq 0$, and
 (iii) $|\epsilon_{e\mu}| \neq 0$ and $|\epsilon_{\mu\tau}| = 0$. We do this for T2HKK and NH. 
 The value of $\chi^2$ for these three cases is 25.46, 17.54 and 24.61, respectively. From these, it is clear that we have a greater effect from $\epsilon_{\mu\tau}$
on the sensitivity than that from $\epsilon_{e \mu}$. We list the values of the different oscillation parameters corresponding to $\chi^2$, as mentioned above in Table \ref{tab1}.
To understand this further, in Fig. \ref{fig:fig2} we plotted the appearance channel probability versus energy for the T2HKK baseline for cases (ii) and (iii) along with the standard, i.e.,
without the NSI. The values of $\theta_{23}$ and $\dcp$ are $45^\circ$ and $-90^\circ$, respectively. The left panel is for neutrinos and the right panel is for antineutrino. In the panels, we also show the
corresponding fluxes (in arbitrary units). From the panels, we can see that within the energy range $E < 2$ GeV (which is the region of interest for T2HKK) the separation between the standard
and green curves is less conspicuous than the separation between the standard and purple curves. This explains why introducing $\epsilon_{\mu\tau}$
affects the sensitivity in a more significant way than introducing $\epsilon_{e\mu}$.

For comparison with the HK atmospheric neutrino observation,
which is analyzed in Refs.\,\cite{Fukasawa:2016nwn}
and \cite{Fukasawa:2017thesis},
in Fig.\,\ref{fig:fig3} we show the excluded regions
at 2$\sigma$ and 3$\sigma$ for T2HKK, DUNE and
HK atmospheric neutrino observations.
The analysis of the HK atmospheric neutrino observation
in Refs.\,\cite{Fukasawa:2016nwn}
and \cite{Fukasawa:2017thesis} was performed using
codes that were applied in Ref.\,\cite{Foot:1998iw,Yasuda:1998mh,Yasuda:2000de,Fukasawa:2015jaa}
under the assumption
that the HK fiducial volumes are 0.56 Mton,
which is the old design of HK, and that the observation is conducted for 12 years,
that the HK detector has the same detection efficiencies as those
of Super-Kamiokande (SK), and the HK atmospheric neutrino data
comprise the sub-GeV, multi-GeV and upward going $\mu$ events as in
the case of SK.
In the case of a normal hierarchy,
we can see from Fig.\,\ref{fig:fig3} that
the sensitivity of the HK atmospheric neutrino experiment
is better than that of the accelerator-based experiments,
particularly with respect to $\epsilon_D$.
This is because the atmospheric neutrino experiment
has information from a wide range of
baseline lengths up to the diameter of the Earth ($\sim$ 13000km)
and it is more sensitive to the matter effect.
By contrast, in the case of an inverted hierarchy,
the sensitivity of the HK atmospheric neutrino experiment
is inferior.  This is because 
atmospheric neutrino experiments with water \cnv\ detectors cannot distinguish neutrinos from antineutrinos and
measure only the sum of neutrinos and antineutrinos.
This leads to a destructive phenomenon in which
the deviations of the neutrino and antineutrino modes
are averaged out\,\cite{Fukasawa:2015jaa}.
In the case of accelerator-based experiments,
which separately measure the neutrino and antineutrino modes,
such a destructive phenomenon does not occur
and the sensitivity for an inverted hierarchy
is almost the same as that for a normal hierarchy.

In Figs.\,\ref{fig:fig1} - \,\ref{fig:fig3} we assume that the true oscillation parameters are
$\theta_{23}=45^o$ and $\dcp=-90^o$.
We also studied the dependence of the excluded regions
on
the true oscillation parameters
$\theta_{23}$ and $\dcp$, and the results are
given in Fig.\,\ref{fig:fig4}.
The first two rows are for T2HKK, whereas the third and fourth rows correspond to DUNE. Each panel of Fig. \ref{fig:fig4} corresponds to a particular true value of 
$\theta_{23}$, and the four contours correspond to four different values of $\delta_{CP}$.
We considered three choices of a true $\theta_{23}$, which are $41^\circ$, $45^\circ$ and $49^\circ$ along with four choices of $\dcp$, which are $0^\circ,~90^\circ,~180^\circ,270^\circ$.
In these plots, the red and black circles are
the best fit points for $f=u$ and $f=d$ from the global
analysis, respectively. From these plots the following features can be observed:
\begin{itemize}
 \item The dependence of the excluded regions
on $\theta_{23}$ is small, whereas
the dependence on $\dcp$ is relatively large.

\item The sensitivity of T2HKK for NH and IH is almost same whereas for DUNE the sensitivities are different in NH and IH. In fact for DUNE, the sensitivity in IH is slightly better than NH.
This can be attributed to the fact that for T2HKK we have taken a 1:3 running of neutrino and antineutrino beam, whereas for DUNE this ratio is 1:1.

\item In T2HKK the sensitivities corresponding to $\dcp = 90^\circ$ and $270^\circ$ are almost same, but this is not the case for DUNE,
which may be due to the fact that T2HKK will use a narrow band flux
and mainly covers the second oscillation maximum, whereas for DUNE the flux is wide band and it covers both the first and second oscillation maximum.

\item Among the four choices of $\dcp$, the sensitivity is poor for $\dcp = 180^\circ$. This is true for both T2HKK and DUNE.

\item The best-fit points can be ruled out at $3\sigma$ for all combinations of $\theta_{23}$ and $\dcp$ in T2HKK and (DUNE, NH) except for, $\dcp = 180^\circ$. For (DUNE, IH), even the best-fit
points for $\dcp=180^\circ$ can be excluded at $3 \sigma$.
\end{itemize}

Finally in Table \ref{tab2}, we give $\chi^2$ for the best-fit point ($\epsilon_D$, $\epsilon_N$) =(-0.14, -0.03) in NH. The numbers in the table also confirm that
the capability of T2HKK and DUNE to exclude the NSI best fit point does not depend much on the true value of $\theta_{23}$ but does heavily depend uon the true value of $\dcp$.
The sensitivity is at maximum for $\dcp = 0^\circ$ and is worst for $\dcp = 180^\circ$ in NH. From the table, we can also understand that the capability of T2HKK to exclude this particular
best-fit point is better than DUNE for $\dcp = 0^\circ$, ($\dcp = 90^\circ$, $\theta_{23}=49^\circ$), and ($\dcp = -90^\circ$, $\theta_{23}=41^\circ$).

\section{Conclusion\label{sec5}}

In this study, we considered the sensitivity
of the future accelerator-based neutrino long-baseline
experiments T2HKK and DUNE
to the NSI, which was suggested based on
the tension between the mass squared differences
from the solar neutrinos and KamLAND data.
We provided the excluded regions
in the ($\epsilon_D$, $\epsilon_N$) plane,
and it turns out that the sensitivity of DUNE
is slightly better than that of T2HKK.
We found that the both experiments
will exclude some of the regions
suggested by the global analysis
for $f=u$ and $f=d$,
although it is difficult for
both experiments to exclude
the region near the standard scenario point.
If there are no non-standard interactions in
nature, then the best-fit point
of the combined analysis 
of the solar neutrino and KamLAND data by 
Ref.\,\cite{Gonzalez-Garcia:2013usa} can be excluded at
more than $10\sigma$~ for $f=u$ and $f=d$,
whereas the best fit point
of the global analysis~ for $f=u$ and $f=d$~
in Ref.\,\cite{Gonzalez-Garcia:2013usa}
can be excluded at $3\sigma$ by T2HKK and DUNE for most of the parameter space.
Although we have discussed only the two cases of $f=u$ and $f=d$, we
expect that the sensitivity of T2HKK and DUNE to the NSI parameters
($\epsilon_N$ in particular) is better than the existing experiments for wide
range of $\eta$ because the $\chi^2$ distribution is
smooth between $\eta=26.6^\circ$ and $\eta=63.4^\circ$ on the
left panel of Fig.4 of Ref.\,\cite{Esteban:2018ppq}, and
the allowed region is expected to more or less similar to
the one which we obtain in this work.
However,
if the NSI exists and the NSI parameter $\eta$ happens to be
close to $\eta_0=-43.6^\circ$, then both
T2HKK and DUNE will have no sensitivity to
the NSI.  In this case, however,
T2HKK and DUNE are expected to give
stronger bound on the deviation $|\eta-\eta_0|$ because the allowed
region by T2HKK and DUNE in our analysis
for $f=u$ and $f=d$ is smaller than that allowed by the
existing experiments.
We found that accelerator-based
long baseline experiments are more sensitive to
the parameter $\epsilon_N$ than to
$\epsilon_D$.
The sensitivity of the two experiments
were demonstrated to be comparable to, or
in the case of an inverted hierarchy,
even better than that
of the HK atmospheric neutrino experiment.

If the tension between
the solar and KamLAND experiments
is due to the NSI in neutrino propagation
and if the true values of the
parameters lie near the best fit point~for $f=u$
or $f=d$,
we may be able to see an affirmative
signal in these long baseline experiments
in the future.

\section*{Acknowledgments}
The authors would like to thank Shinya Fukasawa for his help with the atmospheric neutrino code.
This research was partly supported by a Grant-in-Aid for Scientific
Research of the Ministry of Education, Science and Culture, under
Grant Nos. 25105009, 15K05058, 25105001, and 15K21734.


\bibliography{t2hkk-nsi-solar}
\end{document}